\documentclass[useAMS,usenatbib]{mn2e}

\usepackage{graphicx}
\usepackage{graphics}
\usepackage{epstopdf}
\usepackage{amsmath}
\usepackage{amssymb}
\usepackage{bm}
\usepackage{verbatim}
\usepackage{dsfont}

\newcommand{\Hb}{H$\beta$}

\begin{document}

\title[Modeling reverberation mapping data II]{Modeling reverberation mapping data II: \\
dynamical modeling of the Lick AGN Monitoring Project 2008 dataset}

\author[A. Pancoast et al.]
{Anna Pancoast$^1$,
Brendon J. Brewer$^2$,
Tommaso Treu$^{1, 6}$, 
 Daeseong Park$^{3, 4}$, 
\newauthor
 Aaron J. Barth$^4$,
 Misty C. Bentz$^5$,
 and
 Jong-Hak Woo$^3$ \\
$^1$Department of Physics, University of California, Santa Barbara, CA 93106, USA; pancoast@physics.ucsb.edu \\
$^2$Department of Statistics, The University of Auckland, Private Bag 92019, Auckland 1142, New Zealand \\
$^3$Astronomy Program, Department of Physics and Astronomy, Seoul National University, Seoul 151-742, Republic of Korea \\
$^4$Department of Physics \& Astronomy, 4129 Frederick Reines Hall, University of California, Irvine, CA 92697-4575, USA \\
$^5$Department of Physics and Astronomy, Georgia State University, Atlanta, GA 30303, USA \\
$^6$Current address: Physics and Astronomy Building, 430 Portola Plaza, Box 951547, Los Angeles, CA 90095-1547, USA
}

\maketitle

\begin{abstract}
We present dynamical modeling of the broad line region (BLR) for a
sample of five Seyfert 1 galaxies using reverberation mapping data taken
by the Lick AGN Monitoring Project in 2008.  By
modeling the AGN continuum light curve and H$\beta$ line profiles directly
we are able to constrain the geometry and kinematics of the BLR 
and make a measurement of the black hole mass that does not depend upon the
virial factor, $f$, needed in traditional reverberation mapping analysis.  
We find that the geometry of the BLR
is generally a thick disk viewed close to face-on. 
While the H$\beta$ emission is found to come
preferentially from the far side of the BLR, the mean size
of the BLR is consistent with the lags measured with
cross-correlation analysis. 
The BLR kinematics are found to be consistent with 
either inflowing motions or elliptical orbits,
often with some combination of the two.  We measure black hole masses
of  $\log_{10}(M_{\rm\,BH}/M_\odot)=6.62^{+0.10}_{-0.13}$ for Arp 151,
$7.42^{+0.26}_{-0.27}$ for Mrk 1310, 
$7.51^{+0.23}_{-0.14}$ for NGC 5548, 
$6.42^{+0.24}_{-0.18}$ for NGC 6814, 
and $6.99^{+0.32}_{-0.25}$ for SBS 1116+583A.
The $f$ factors measured individually for each AGN are found to correlate with inclination angle, 
although not with $M_{\rm\,BH}$, $L_{5100}$, or FWHM/$\sigma$ of the emission line profile.
\end{abstract}


\section{Introduction \label{sect_intro}}
While active galactic nuclei (AGNs) are thought to be powered by accretion onto
supermassive black holes at the centers of most galaxies, the geometry and
dynamics of the surrounding regions are not well understood.  In the standard
model of AGNs \citep{antonucci93, urry95}, the region directly outside the
accretion disk is the broad line region (BLR), where broad line emitting gas
moves at velocities of $10^{3-4}$ km\,$\rm{s^{-1}}$ within the Keplerian
potential of the black hole.  Measurements of the distance of this gas from the
central ionizing source in the accretion disk are on the order of light days
for lower luminosity AGNs such as Seyfert galaxies and this distance increases
with AGN luminosity \citep{wandel99, kaspi00, bentz06, bentz13}. 

The geometry and dynamics of the BLR can be further constrained by reverberation
mapping measurements, where changes in the AGN continuum emission are monitored
alongside the echo of these same changes in the BLR emission lines
\citep{blandford82, peterson93, peterson04}.  The time lag $\tau$ between
changes in the AGN continuum flux and those of the broad emission
lines is interpreted as a measure of the average radius of the BLR and
traditionally measured using the cross-correlation function.  The time
lag can then be combined with BLR gas velocities $v$ taken from the
width of the broad emission line to measure a virial product that has the dimensions of black hole mass.  
The virial product $M_{\rm vir} = c\tau v^2/G$ is
related to the true black hole mass $M_{\rm BH}$ by a dimensionless
virial factor $f$ of order unity that is calibrated by aligning the
$M_{\rm BH}-\sigma_*$ relations for active and inactive galaxies
\citep{onken04, collin06, woo10, greene10b, graham11, park12b, woo13, grier13b}.  Currently, the
uncertainty in mean $f$ of $\sim 0.4$ dex is the largest uncertainty in reverberation
mapped black hole masses \citep[e.g.][]{park12b}.  Since the sample of $\sim 50$
reverberation mapped AGNs is responsible for calibrating single-epoch $M_{\rm BH}$
estimates applied to much larger samples of AGNs through the
BLR-size-to-luminosity relation \citep{vestergaard06, mcgill08, vestergaard11},
it is important to measure $M_{\rm BH}$ in reverberation mapped AGNs with as
few assumptions and added uncertainties as possible.

There is more information in high quality reverberation mapping data
than a single estimate of the time lag.  This is illustrated by analysis
of high-quality datasets that show clear velocity-resolved lag
structure across the emission line profile in a number of AGNs.  Many
of these velocity-resolved lag estimates are consistent with bound
orbits in a Keplerian potential \citep{bentz09, denney10, barth11a, barth11, grier13a},
but some show additional kinematic signatures consistent with
inflowing or outflowing gas
\citep{bentz09, denney10, grier13a}, where non-gravitational forces may be at work.  Much recent work has focused on trying to
recover this additional information about the BLR geometry and dynamics through
more sophisticated analysis techniques.  One method for doing this is to recover
the transfer function, which is the distribution of time lags either for the
integrated emission line flux or as a function of line-of-sight velocity
\citep[see][and references therein]{bentz10, grier13a}.  This is the approach
used in the code MEMECHO, which recovers the transfer function using a flexible
parameterization for the transfer function shape and prevents over-fitting by
asserting that the solution has a high entropy \citep{horne91, horne94}.  The main
advantage to recovering the transfer function is that it does not require
adopting any specific model for the geometry and dynamics of the BLR
\citep[see][for an application of a pixelated method to the
reconstruction of the transfer function]{krolik95}.  
This also means that in order to relate features in the recovered transfer function to
geometrical and dynamical properties of the BLR at a quantitative level, models must eventually be
constructed and compared to the transfer function. 

Members of our team have been working on an alternative approach that
involves modeling reverberation mapping data directly to constrain the
geometry and dynamics of the BLR \citep[see][]{pancoast11, brewer11,
pancoast12}.  While this method requires adopting a specific BLR
model, the direct modeling approach is quite general and allows for
the use of any BLR model whose consequences are fast to compute.  The
direct modeling method can then be used to determine which theoretical
models of the BLR geometry and dynamics are preferred by the data.  By
formulating the method as a problem of Bayesian inference, parameter
estimation (within the context of a particular BLR model), and model
selection (comparison of distinct BLR models) are both possible.  In
addition to constraining the geometry and dynamics of the broad line
emitting gas, the direct modeling approach allows for independent
measurement of the black hole mass without relying on the normalizing
factor $f$ required by the traditional analysis.  This means that with
high-quality reverberation mapping datasets, the uncertainty in black
hole masses from dynamical modeling could be substantially less than
the $\sim 0.4$ dex introduced by assuming a mean value for $f$
\citep{pancoast11}.  Finally, independent estimates of the black hole
mass can be compared to $M_{\rm vir}$ to obtain measurements of $f$
for individual AGNs.  With a large enough sample of AGNs with
individual $f$ values from direct modeling, we have another method of
calculating a mean $f$ factor for different AGN populations.

The direct modeling method works by using the continuum flux light curve and a
model of the geometry and dynamics of the BLR to create a time series of simulated
broad emission line profiles.  The simulated line profiles can then be directly
compared to the reverberation mapping data line profiles to infer which BLR model
parameter values best reproduce the data.  Working with the formalism of
Bayesian inference, we calculate the joint posterior probability distribution function (PDF) for the model
parameters. We have focused on
developing a simply parametrized phenomenological model of the BLR to map the
density of broad line emission in position and velocity space.  We have
purposefully excluded any constraints from photoionization physics or radiative
transfer at this stage, because doing so requires assuming a relation between
the distribution of broad line flux emission and the distribution and density
of the emitting gas.  While extending the model to infer the physics of the BLR
gas is certainly of interest, we have found that a flexible simply parameterized
model is sufficient to reproduce the line profile shape and variability
characteristics in current reverberation mapping spectral datasets as
demonstrated for Arp 151 and Mrk 50 \citep[e.g.][]{brewer11, pancoast12}.  

In addition to a model of the BLR, it is necessary to include a model for the
AGN continuum light curve, since it is necessary to evaluate the AGN continuum
flux at arbitrary times to produce simulated broad emission line profiles.  We
use Gaussian processes to interpolate between the continuum light curve
datapoints, which allows us to include the uncertainty from the interpolation
process into the final uncertainty on the BLR model parameters.  The simplified
version of Gaussian processes that we use is the same as a continuous time
first-order autoregressive process (CAR(1)), which has been found to be a good
model for larger samples of AGN light curves
\citep{kelly09, kozlowski10, macleod10, zu11, zu12}.  This model for AGN
continuum variability has been used in a number of other reverberation mapping
analyses.  For example, \citet{zu11} use a CAR(1) process model for the AGN
continuum light curve interpolation in their code JAVELIN to measure the time
lag between the continuum and an integrated emission line light curve using a
top-hat transfer function.  Similarly, \citet{li13} use a CAR(1)
process model in their code to directly model integrated emission line
reverberation mapping data based on the BLR model of \citet{pancoast11}.

In this paper we apply the direct modeling method to the Lick AGN Monitoring
Project (LAMP) 2008 reverberation mapping dataset 
\citep{walsh09, bentz09}.\footnote{The LAMP 2008 spectroscopic dataset is available for download here: http://www.physics.uci.edu/$\sim$barth/lamp.html}  
The LAMP 2008 campaign observed 13 AGNs using spectroscopy from the Shane
Telescope at Lick Observatory and Johnson {\it V} and {\it B} broad-band photometry from a
number of ground-based telescopes.  
We focused our direct modeling on the \Hb\ line of the 9 objects with measurable
time lags, using the broad and narrow \Hb\ emission line components isolated from the stellar continuum and Fe\,{\sc ii} lines using 
spectral decomposition techniques by \citet{park12a}.
 Out of the 9 objects to which we applied our direct modeling method, only 5
objects showed sufficient continuum and line variability to allow for constraints
on the geometry and dynamics of the BLR.  Of the five objects with successful
direct modeling of the \Hb\ line presented here, one of the objects, Arp 151,
has previous direct modeling results as described by \citet{brewer11}.  There
are two main differences between the direct modeling of \citet{brewer11} and
this work: the first is that we use the spectral decompositions from \citet{park12a} instead
of \citet{bentz09}, and the second is that the model of the BLR has since
been substantially improved.  
Improvements to the BLR model include a new dynamics model
and two additional geometry model parameters that add flexibility
to the shape of the BLR.  In addition, we now model the narrow emission
line component of \Hb\ using the width of the [\mbox{O\,{\sc iii}}]$\lambda5007$ narrow emission line
and calculate the instrumental resolution for each epoch of spectroscopy separately.

Our focus in this paper is to apply the direct modeling method to the
remainder of the LAMP 2008 sample, including reanalysis of Arp 151.  In
Section~\ref{sect_data} we describe the LAMP 2008 data used in our analysis.  
In Section~\ref{sect_model} we briefly review our model for the BLR with further
details to be found in a companion paper \citep[][hereafter paper I]{pancoast14}.  
In Section~\ref{sect_results} we present the results of
our analysis for the five successfully modeled AGNs in the LAMP 2008 sample.  
Finally, in Section~\ref{sect_conclusions} we summarize our results and 
discuss their implications for future direct modeling work.  All quantities related to properties of
the BLR are given in the rest frame of the system.

\section{Data}
\label{sect_data}
Our sample of AGNs was observed in the LAMP reverberation mapping campaign in
2008.  The first part of the data consists of Johnson {\it B} and {\it V} broad-band AGN continuum
light curves measured using standard aperture photometry techniques, as described
by \citet{walsh09}.  The  {\it B} and {\it V} band images were taken at a number of
telescopes, including the 30-inch Katzman Automatic Imaging Telescope (KAIT),
the 2-m Multicolor Active Galactic Nuclei Monitoring telescope, the Palomar
60-inch telescope, and the 32-inch Tenagra II telescope.  For direct modeling
of each AGN, we choose to use either the  {\it B} or {\it V} band light curve depending on
which has more data points, better sampling of variability features, and higher
overall variability.  In general, the choice of {\it B} or {\it V}-band AGN continuum light
curve does not change our results.  

The second part of the data comprises light curves of broad and narrow \Hb\ line
profiles.  Measurement of the \Hb\ line profiles was
done in two ways: \citet{bentz09} isolated the \Hb\ flux by fitting a local
linear continuum underneath the \Hb\ and O\,{\sc iii} lines, while
\citet{park12a} applied a multi-component fit to isolate the \Hb\ line from the
AGN continuum, stellar continuum, and Fe\,{\sc ii} emission lines.  Due to the
non-negligible contribution of the stellar continuum and Fe emission lines to
the five LAMP 2008 objects considered here, we
performed direct modeling on the \Hb\ emission line profiles as measured by
\citet{park12a}.  The final spectra we use here for modeling include the broad and narrow
\Hb\ line profiles, as well as the spectral decomposition residuals, equivalent
to subtracting off all other components from the original spectrum.  The details
of the final spectra, including wavelength range used for direct
modeling, are given in Table~\ref{table_data}.

One other important parameter of the spectral dataset is the instrumental
resolution, which is used to smooth the simulated emission line profiles.  The
instrumental resolution was measured by \citet[][see their Table 11]{bentz09}
for four of the five objects by comparing the [\mbox{O\,{\sc iii}}]$\lambda5007$ line
widths to the values measured by \citet{whittle92}.  However, there were
variations in the [\mbox{O\,{\sc iii}}]$\lambda5007$ line width over the duration
of the reverberation mapping campaign due to small changes in the observing and instrumental conditions.
For this reason, we calculate the instrumental resolution, $\Delta\lambda_{\rm dis}$, for each night independently
using the width of the [\mbox{O\,{\sc iii}}]$\lambda5007$ line from 
the spectral decomposition by \citet{park12a}, $\Delta\lambda_{\rm obs}$, and the intrinsic line width
as measured by \citet{whittle92}, $\Delta\lambda_{\rm true}$, by subtracting them in quadrature:
\begin{equation}
\Delta\lambda^2_{\rm dis} \approx \Delta\lambda^2_{\rm obs} - \Delta\lambda^2_{\rm true}.
\end{equation}
In order to include the uncertainties in these line width
measurements, we consider both the measured and intrinsic widths of the 
[\mbox{O\,{\sc iii}}]$\lambda5007$ line to be free parameters with Gaussian priors 
centered on the measured values and with dispersions given by the quoted measurement
uncertainties.  For the one object, SBS
1116+583A, without a comparison line width by \citet{whittle92}, we use a value in the
middle of the range of the values of the other four objects.      
The values of the intrinsic [\mbox{O\,{\sc iii}}]$\lambda5007$ line width used in this analysis
 are given in the last column of
Table~\ref{table_data} as the line dispersion in \AA, converted from the FWHM of the 
line widths in units of km\,s$^{-1}$ listed in \citet{whittle92} assuming the Gaussian conversion of 2.35, for all objects except SBS 1116+583A.

\begin{table*}
 \begin{minipage}{155mm}
  \caption{Properties of the LAMP 2008 spectra and photometry.  Band is
the Johnson broad-band filter.  No. Continuum Epochs is the number of data
points in the AGN continuum light curve in the band given by column 2.  No.
Line Epochs is the number of spectra in the broad emission line time series.  
No. Spectral Pixels is the number of pixels in the \Hb\ spectrum between the
wavelength ranges given in the next column. Wavelengths are in the rest frame. 
Intrinsic [\mbox{O\,{\sc iii}}]$\lambda5007$ Width is the intrinsic line dispersion $\sigma$ of the narrow
[\mbox{O\,{\sc iii}}]$\lambda5007$ emission line used for calculating the instrumental resolution.}
  \label{table_data}
  \begin{tabular}{@{}ccccccc}
   \hline
   Object & Band  &  No. Continuum &  No. Line & No. Spectral  & Wavelength Range &  Intrinsic [\mbox{O\,{\sc iii}}]$\lambda5007$  \\
              &            &  Epochs             &  Epochs  & Pixels            & (\AA)                       &    Width (\AA) \\
   \hline
   Arp 151                   & {\it B}   &  84  & 43  & 73   & $4792.3 - 4933.4$   & $1.562\pm0.071^a$ \\
   Mrk 1310                & {\it B}   &   50  & 47  & 51   &  $4815.5 - 4913.6$   & $0.852\pm0.071^a$ \\
   NGC 5548              & {\it V}   &   57  & 51  & 171  &  $4706.5 - 5040.9$   & $2.910\pm0.071^a$ \\
   NGC 6814              & {\it V}   &   46  & 45  & 81    &  $4776.7 - 4935.8$    & $0.888\pm0.071^a$ \\
   SBS 1116+583A     &{\it B}   &   56  & 50   & 67   &  $4797.3 - 4925.7$    & $1.4\pm 0.3$ \\
   \hline
  \end{tabular}
  \medskip
  $^a$ These values are converted from measurements by \citet{whittle92} assuming an uncertainty of $10$\,km/s.
  \end{minipage}
  \end{table*}

\section{The Dynamical Model of the Broad Line Region}
\label{sect_model}
In this section we give a brief overview of our simply parameterized
phenomenological model of the BLR geometry and dynamics, with full model
details given in paper I.   
We model the distribution of broad line flux emission
by the density of many point particles that instantaneously and linearly reprocess the AGN continuum flux and reemit some fraction of it back towards 
the observer with time lags that depend upon the point particles' positions.  
The velocities of the point particles then
determine how redshifted or blueshifted the broad line flux from the point
particles is relative to the systemic velocity of the BLR system.  This means
that in addition to a model describing the distribution of point particles in
position and velocity space, we must also model the AGN continuum flux in order
to evaluate it at arbitrary times.   

For our model of the AGN continuum light curve we use Gaussian processes, which
allows us to sample the AGN continuum variability on scales much smaller than
the typical one day cadence between datapoints.  This AGN continuum variability
model allows us to include the uncertainty from interpolation in our final
uncertainties in the BLR model parameters, as well as allowing us to extrapolate
beyond the ends of the light curve in order to evaluate long time
lags \citep[for an illustration see][]{pancoast11}.  

We model the BLR geometry by defining the physical distribution of point
particles.  The radial distribution is given by a shifted Gamma distribution,
which can reproduce Gaussian, exponential or heavy-tailed radial distributions
depending on the value of its shape parameter.  The point particles
are also constrained to be within an (unknown) opening angle, which allows for
spherical geometries ranging to thin disk geometries.  The BLR is then viewed
by an observer with an inclination angle ranging from face-on to edge-on.  
Finally, there are a number of non-axisymmetric effects that allow for more
flexibility in the BLR geometry.   
These include preferential emission of the point particles from the 
near or far side of the BLR along the observer's line of sight,
a transparent to opaque mid-plane, and the possibility of increased
emission from the edges of the BLR disk, relative to the inner portion.

Similarly, we model the BLR dynamics by defining the velocity distribution of
point particles as a function of position and black hole mass.
We draw the point particles' velocities from distributions in
the space of radial and tangential velocities, centered around the circular
orbit values or from around the radial escape velocity values for inflowing or
outflowing orbits.  We allow for a combination of elliptical orbits centered
around the circular orbit values plus either inflow or outflow centered around
their respective radial escape velocities.  To allow for mostly bound inflowing
or outflowing orbits, we also allow the distributions of inflowing and outflowing
orbits to be centered anywhere along an ellipse between the radial escape
velocity and the circular orbit velocity.  In order to add more flexibility, we
also allow for additional macroturbulent velocities.  Finally, we include
gravitational redshift and the full expression for doppler shift when moving
between the velocities of the point particles and the wavelengths of broad line
flux emission.  The exact definitions of the geometry and dynamics model
parameters are given in Appendix~\ref{appendix_params}, with more detailed
descriptions given in paper I.

In addition to modeling the broad emission line component using a model
for the BLR, we also model the narrow emission line component using the 
width of the [\mbox{O\,{\sc iii}}]$\lambda5007$ narrow emission line.  Since
the width of the [\mbox{O\,{\sc iii}}]$\lambda5007$ line is a combination of
intrinsic line width and instrumental resolution, we use measurements of the
intrinsic line width to constrain the instrumental resolution for smoothing of the
broad emission line component.

We explore the parameter space of the BLR model and AGN continuum variability
model using Diffusive Nested Sampling \citep{brewer11b}. This algorithm samples
the posterior distribution for the parameters, and provides the ``evidence''
value which can be used to compare distinct models.
We use a Gaussian likelihood function that
compares the time series of broad emission line profiles of the data to the
simulated versions generated by our BLR model.  Since the model is of finite
flexibility and the spectral data have
high signal to noise and a large number of data points,
it is necessary to soften the likelihood function.  We do this
by defining a temperature $T$ by which the log of the likelihood is divided,
where $T \ge1$.  Temperatures greater than one can be thought of as an
additional source of uncertainty in the likelihood due to the model not
providing a perfect fit to the data, either because the error bars of the data
are underestimated or because the model does not contain enough complexity to
reproduce all features of the data. The use of a temperature can be thought of
as a computationally inexpensive approximation to a correlated noise model.
Since we use Nested Sampling, the choice of temperature $T$ can be made
in post-processing and does not require multiple runs.

\section{Results}
\label{sect_results}
\begin{figure}
\begin{center}
\includegraphics[scale=0.56]{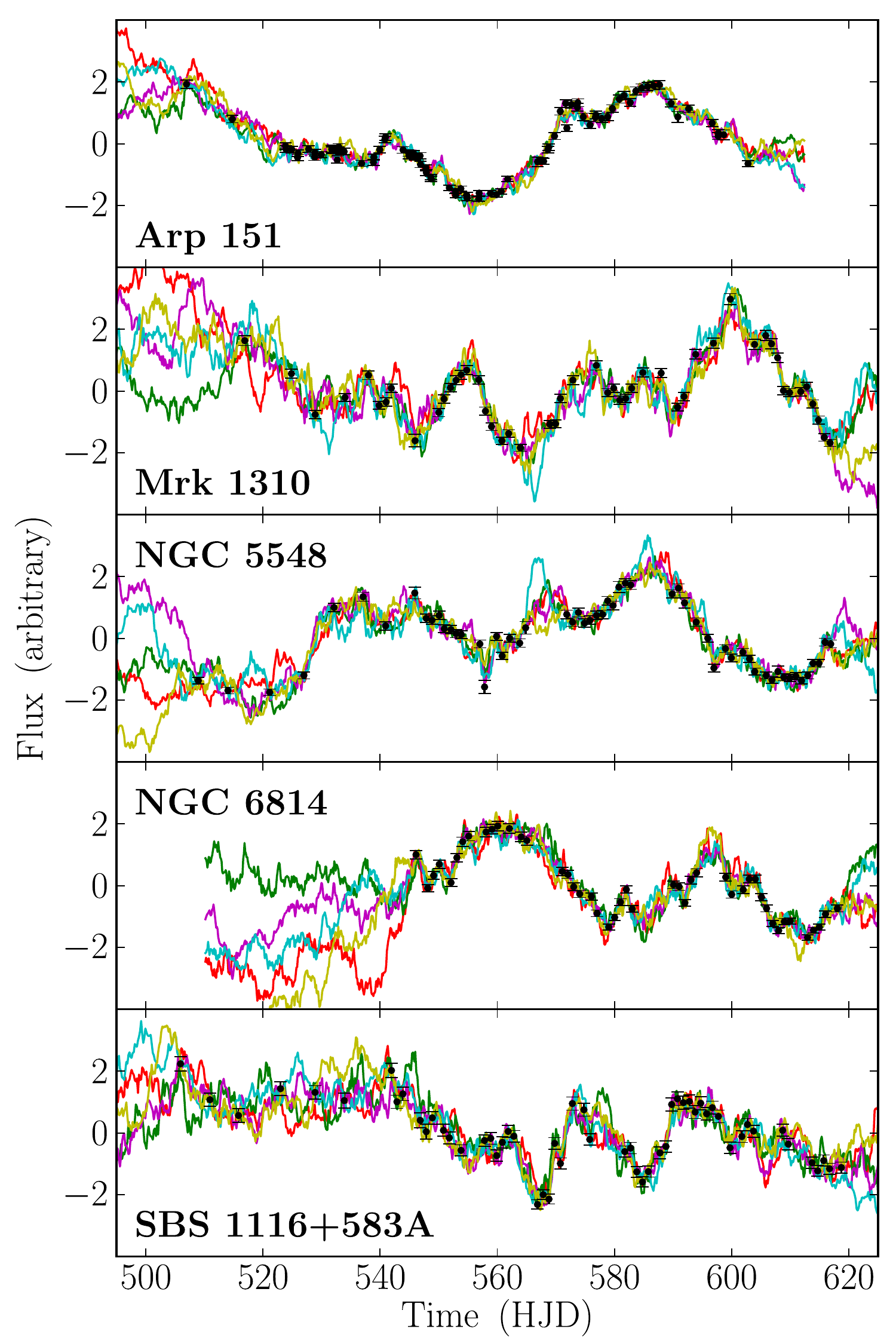}
\caption{ AGN continuum light curves for the five objects in our sample.
The data are shown by black points with error bars and the Gaussian process interpolations
drawn from the posterior PDF are shown by the colored lines.
Top to bottom: Arp 151, Mrk 1310, NGC 5548, NGC 6814, and SBS 1116+583A.} 
\label{fig_cont}
\end{center}
\end{figure}

We now present the results of direct modeling of five AGNs in the LAMP 2008
sample, including Arp 151, Mrk 1310, NGC 5548, NGC 6814, and SBS 1116+583A.  
First, the modeling results are presented in detail for each AGN, including key
posterior PDFs and correlations, the model fit to the data, and the transfer
function.    
Examples of the inferred geometry of the BLR for each AGN are also shown in Figure~\ref{fig_geo}
and the AGN continuum light curves with Gaussian process interpolations drawn
from the posterior are shown in Figure~\ref{fig_cont}.
The posterior median and central 68\% credible intervals for the main BLR model
parameters are given in Table~\ref{table_results}.  Second, we present an
overview of the BLR modeling results to emphasize the similarities and
differences in the sample.  Finally, we calculate the mean $f$ factor for this
sample of five AGNs.

\subsection{Individual Modeling Results}
\subsubsection{Arp 151 (Mrk 40)}
\label{sect_arp151}

\begin{figure}
\begin{center}
\includegraphics[scale=0.44]{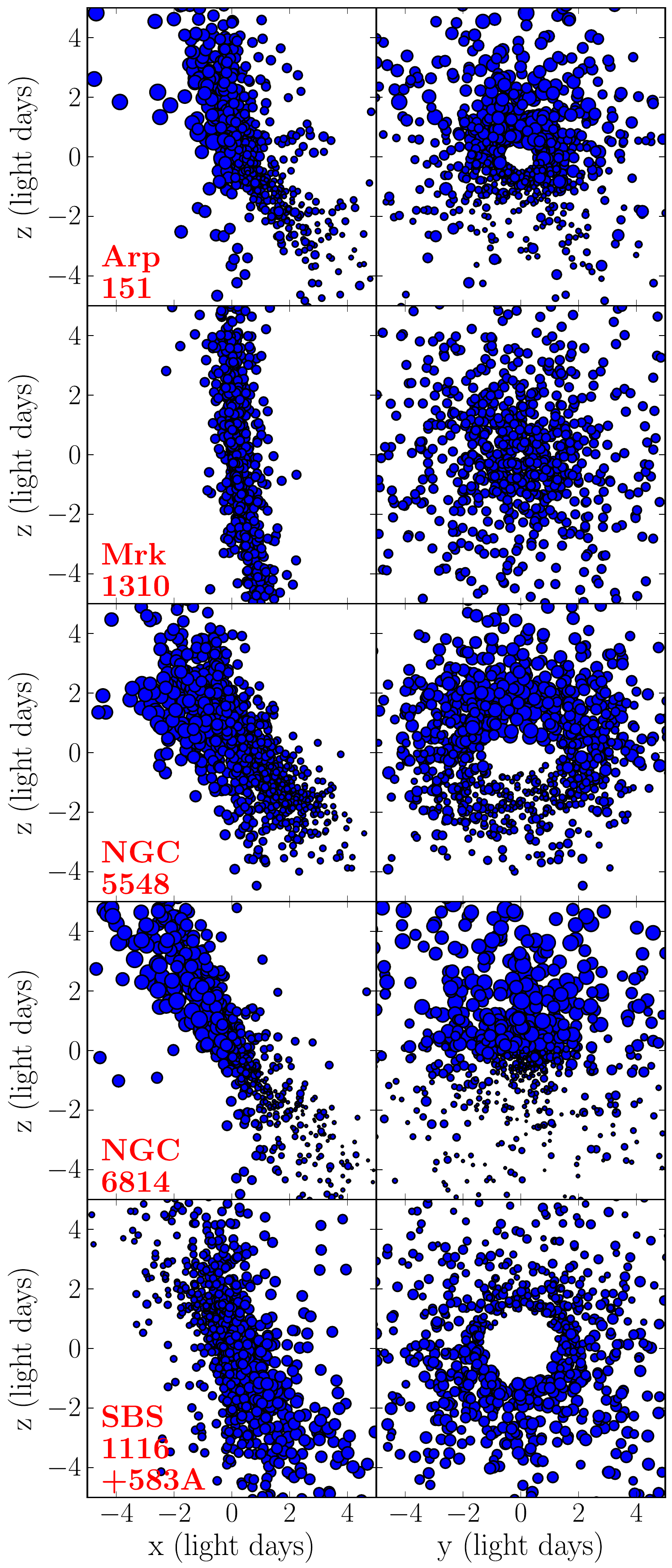}
\caption{Geometries of the BLR for the five objects in our sample.
The left panels show the BLR from along the $y$ axis (the edge-on view), while the right panels 
show the BLR from along the positive $x$ axis (the observer's point of view).
Top to bottom: Arp 151, Mrk 1310, NGC 5548, NGC 6814, and SBS 1116+583A.
Each point corresponds to a point particle in our BLR model and the 
size of the points is proportional to the relative amount of \Hb\ emission coming from each
point particle, given the same incident continuum flux.} 
\label{fig_geo}
\end{center}
\end{figure}

\begin{figure}
\begin{center}
\includegraphics[scale=0.46]{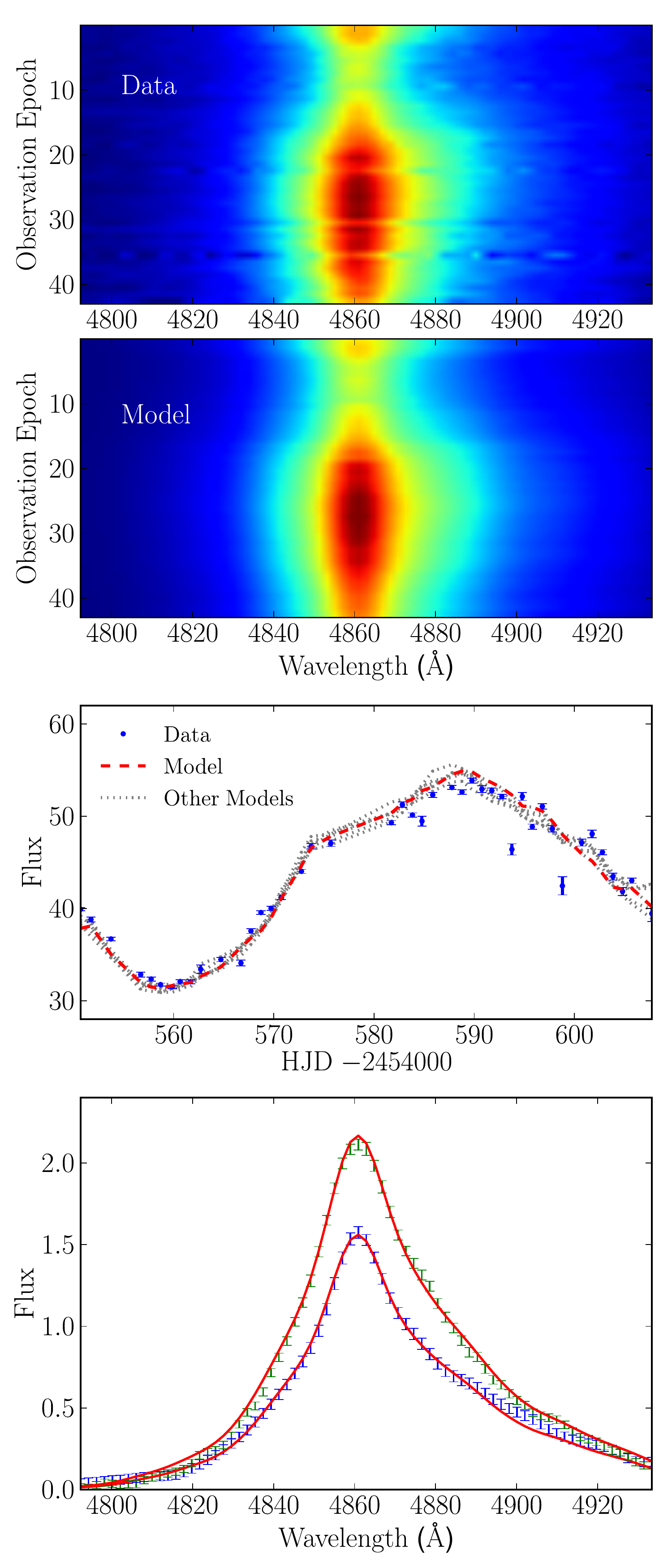}
\caption{Model fit to the broad and narrow \Hb\ line dataset for Arp 151.  Top panel: the \Hb\ spectral
time series of data from \citet{park12a}.  Top middle panel: an example of a simulated \Hb\ spectral
time series for a model drawn randomly from the posterior PDF.  Bottom middle
panel: the integrated \Hb\ line light curve with data from \citet{park12a} given by the blue
points with error bars, the model in the top middle panel shown with the red
dashed line, and additional models drawn from the posterior shown with
the dotted grey lines.  Bottom panel: two examples of the \Hb\ line profile
shown with blue and green error bars with the model fits over-plotted with red
lines.} 
\label{fig_arp151_rainbow}
\end{center}
\end{figure}

\begin{figure}
\begin{center}
\includegraphics[scale=0.35]{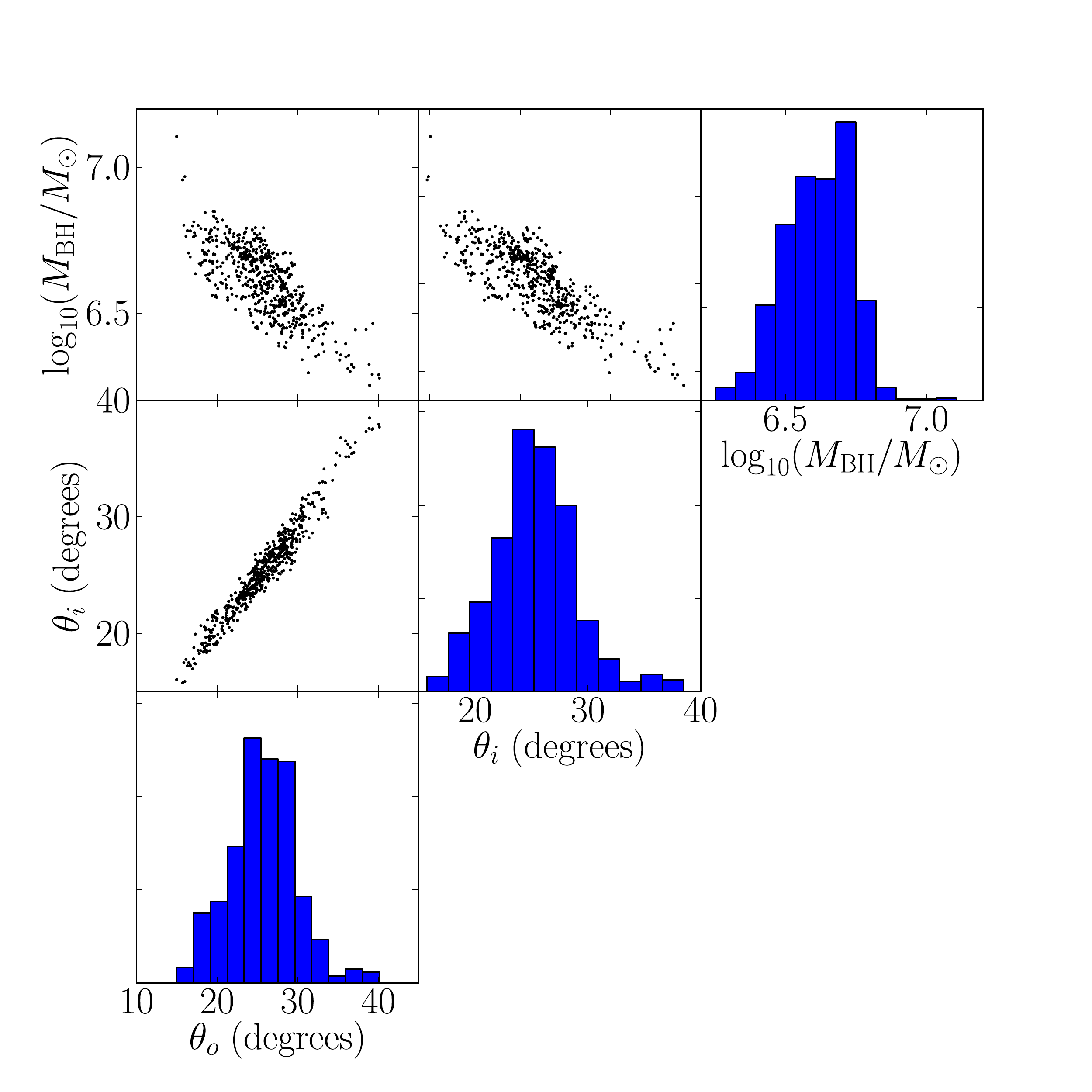}
\caption{Marginal posterior PDFs and correlations between parameters for Arp
151, including black hole mass ($M_{\rm BH}$), inclination angle ($\theta_i$),
and opening angle ($\theta_o$).} 
\label{fig_arp151_cp}
\end{center}
\end{figure}

\begin{figure}
\begin{center}
\includegraphics[scale=0.46]{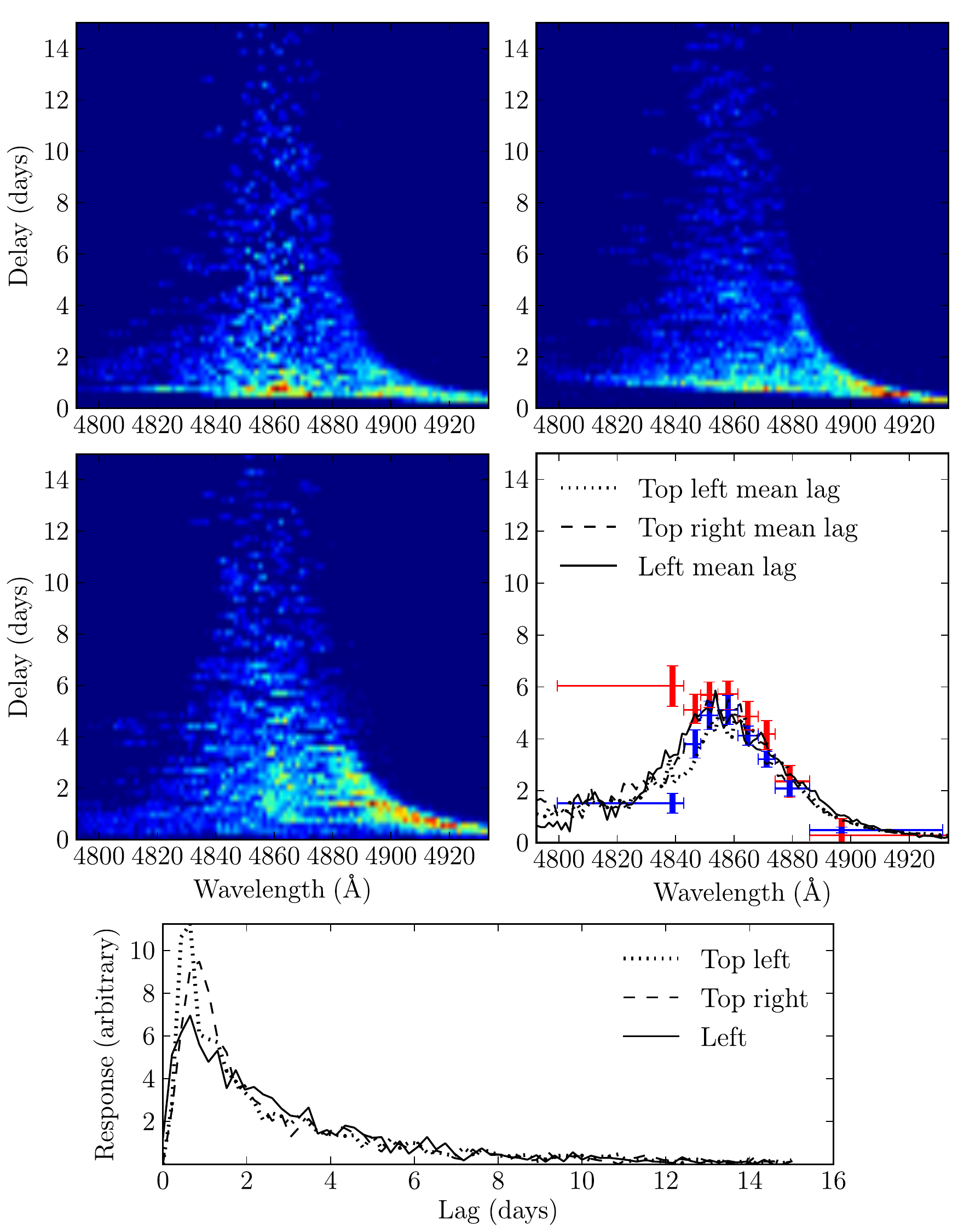}
\caption{Transfer functions for Arp 151.  The top two panels and the middle left
panel are all examples of transfer functions drawn from the posterior PDF of the
model fit.  The middle right panel shows the mean lag as a function of velocity
for each of the three transfer functions from the posterior.  Also shown in this panel
are the velocity-resolved cross-correlation lag measurements from \citet{bentz09} as
red error bars, where the horizontal error bars show the wavelength ranges used in the
integrated light curves.  Our mean lag values in these same bins are shown by the
blue error bars, except for the longest wavelength bin which does not extend
as far to the red as the one by \citet{bentz09}.  The bottom panel
shows the velocity-integrated transfer function for each of the three transfer
functions from the posterior.} 
\label{fig_arp151_tf}
\end{center}
\end{figure}

Both the AGN continuum and broad \Hb\ line showed strong variability over the
duration of the LAMP 2008 campaign, leading to the clearest velocity-resolved
lag measurements of the LAMP 2008 sample \citep{bentz09} and the most detailed
transfer function recovered at the time using MEMECHO \citep{bentz10}.  It is
therefore unsurprising that the direct modeling results for Arp 151 also have
the highest quality of the LAMP 2008 sample.  

Comparison of the spectral time series and time series of simulated spectra, as
illustrated in Figure~\ref{fig_arp151_rainbow}, suggests that the model is able
to fit the overall variability structure of the \Hb\ line profile very well.  
In addition, the integrated model \Hb\ emission line and individual model
spectra show excellent agreement.  The model is unable to capture very short
timescale variations that are either due to noise or processes with much faster
response times than the overall variability of the BLR would suggest.  
Fortunately, such short timescale variations are infrequent and do not appear
to substantially affect inference of the model parameters.

The geometry of the BLR in Arp 151 as traced by \Hb\ emission is inferred to be
a wide thick disk, inclined by $\theta_i = 25.2^{+3.3}_{-3.4}$ degrees relative to the observer ($0=$ face-on).  
The radial distribution of \Hb\ emission has heavier tails than an exponential profile, with
a Gamma distribution shape parameter of $\beta = 1.25^{+0.15}_{-0.16}$, mean radius $r_{\rm mean} = 3.44^{+0.26}_{-0.24}$ light days, and dispersion or radial
width of $\sigma_r = 3.72^{+0.45}_{-0.43}$ light days.  The radial distribution is offset from the origin,
the source of the ionizing photons and visible continuum emission, by $r_{\rm min} = 0.44^{+0.13}_{-0.20}$
light days.  The mean radius equals to within the uncertainties the mean lag of $\tau_{\rm mean} = 3.07^{+0.25}_{-0.20}$ days, which in turn is consistent
with the cross-correlation measured central lag of $\tau_{\rm cent} = 3.99^{+0.49}_{-0.68}$ days by 
\citet{bentz09} to within the uncertainties.  Due to the heavy tails of the radial profile, the median lag of
$\tau_{\rm median} = 1.75^{+0.28}_{-0.23}$ days is significantly shorter.  The opening angle of the disk is well
constrained to be $\theta_o = 25.6^{+3.7}_{-4.0}$ degrees, however more emission is found to
come from the outer faces of the disk ($\gamma = 4.27^{+0.54}_{-0.80}$), making the geometry
closer to a cone.  There is also preferential emission from the far side of
the BLR from the observer ($\kappa = -0.36^{+0.08}_{-0.08}$) and
the mid-plane of the BLR disk is found to be mostly opaque ($\xi = 0.09^{+0.08}_{-0.05}$).
An example of the BLR geometry in Arp 151 for a set of model parameters drawn from the posterior is shown
in Figure~\ref{fig_geo}.

The dynamics of the BLR in Arp 151 are inferred to be dominated by inflowing
orbits, with the fraction of point particles in elliptical orbits only $f_{\rm ellip} =  0.06^{+0.09}_{-0.05}$, or $1-15\%$.
The majority of the point particles are in inflowing orbits as given by the inflow/outflow parameter $f_{\rm flow} = 0.24^{+0.20}_{-0.17}$,
where values of $f_{\rm flow}$ between 0 and 0.5 indicate inflow and values between 0.5 and 1 indicate outflow. 
Comparing the probability for values of $f_{\rm flow}$ between 0 and 0.5 with
the probability for values between 0.5 and 1 indicates a 100\% preference for inflow compared to outflow.
Furthermore, the inferred inflowing orbits are not strictly radial or drawn from a velocity distribution centered on the radial escape velocity, but can be
distributed closer to the circular orbit value, leading to more bound inflowing orbits.  The value of $\theta_e = 12.0^{+10.7}_{-8.3}$
that we infer for Arp 151 indicates that the inflow orbit velocity distribution is rotated about a seventh of the way towards the circular-orbit-centered distribution 
and that more than half of the inflowing orbits are bound.  Finally, we find a negligible contribution to the dynamics of the BLR from macroturbulent velocities,
with the dispersion of additional macroturbulent velocities drawn from a Gaussian distribution of only $\sigma_{\rm turb} = 0.008^{+0.028}_{-0.007}$ times
the circular orbit velocity.

We measure a black hole mass for Arp 151 of $\log_{10}(M_{\rm BH}/M_\odot) = 6.62^{+0.10}_{-0.13}$.  As illustrated in Figure~\ref{fig_arp151_cp}, there
is strong degeneracy between the black hole mass, inclination angle, and opening angle, preventing us from measuring the black hole mass with greater precision.  
The correlation between these parameters is easy to understand if one considers that the BLR model parameters are constrained such that the line-of-sight velocity
matches the width of the emission line: for thin disks, the more face-on the BLR, the higher the black hole mass must be to produce the same line-of-sight velocities. 
The opening angle is also strongly correlated, since a thicker disk allows for larger line-of-sight velocities for a given black hole mass.  

With an independent measurement
of the black hole mass we can use the virial product $M_{\rm vir}$ from traditional reverberation mapping analysis to measure the $f$ factor for Arp 151.  
We use the time lags $\tau_{\rm cent}$ from cross-correlation analysis from \citet{bentz09} 
and measurements of the \Hb\ line width after spectral decomposition from \citet{park12a} to construct two sets of virial products.  
The first type of virial product uses the line dispersion measured from the RMS line profile as the \Hb\ line width and the second uses the FWHM of the mean line profile as the \Hb\ line width.
Values of the $f$ factor calculated using the first type of virial product will be referred to as $f_\sigma$, while values calculated using the second type of virial product
will be referred to as $f_{\rm FWHM}$.  
We obtain the distribution of $f$ for each AGN by subtracting the virial product from the normalized posterior PDF of black hole mass.  
The inferred $f$ factors for the five AGNs in our sample are listed in Table~\ref{table_f}.  
For Arp 151, we measure $f$ factors of $\log_{10}(f_\sigma) = 0.51^{+0.10}_{-0.13}$ and $\log_{10}(f_{\rm FWHM}) = -0.24^{+0.10}_{-0.13}$.

Previous direct modeling results for Arp 151 constrained the black hole mass to
be $10^{6.51 \pm 0.28}\,M_\odot$ and the geometry to be a wide thick disk with
an opening angle of $\theta_o = 34.5^{+10.7}_{-8.6}$ degrees, inclined with respect to the
observer by $\theta_i = 22.2 \pm 7.8$ degrees \citep{brewer11}.  Our improved
modeling results for Arp 151 are completely consistent to within the uncertainties
with these previous modeling results, and clarify the previous ambiguity in
whether the dynamics of \Hb\ in Arp 151 are dominated by inflow or outflow.  

For comparison with work focused on recovering the velocity-resolved transfer
function, we show three transfer functions created from models
drawn randomly from the multi-dimensional posterior PDF in Figure~\ref{fig_arp151_tf}.
While the three transfer functions show slightly different detailed structure, the mean lag as
a function of velocity is very similar for all three, as is the velocity-integrated transfer function.  
In addition, all three velocity-resolved transfer functions show at least some preference for prompt response on the red side of the line 
profile, indicative of inflow kinematics.  The same inflow signatures were found in the transfer function recovered using MEMECHO \citep{bentz10},
as well as the velocity-resolved lag measurements, shown in red in the middle right panel of Figure~\ref{fig_arp151_tf}.  
The discrepancy in the blue wing of the line between the velocity-resolved lag measurements from CCF analysis (in red in Figure~\ref{fig_arp151_tf}) and
dynamical modeling (in blue) is due to a combination of data preprocessing and systematics from measurement of the time lag.  
Recalculating the velocity-resolved time lags from CCF analysis using the same datasets as in the dynamical modeling
decreased the discrepancy in the bluest lag bin by $\sim 1.5$ days while remaining consistent with the values from \citet{bentz09}.  The remaining 
discrepancy is due to the difference between the true mean time delay and the time delay proxy estimated by CCF analysis.  
We confirm this by creating
velocity-resolved light curves using the inferred models of the BLR for Arp 151, calculating and showing that the CCF time lag from those model 
light curves and the time lags from dynamical modeling are consistent with the values from \citet{bentz09}.
However there are residual differences between the transfer functions from direct modeling and MEMECHO, including response
in the blue wing of the line, where direct modeling finds significantly shorter lags, and prompt response of the 
line emission at line center, where the MEMECHO solution finds no prompt response.

\subsubsection{Mrk 1310}

\begin{figure}
\begin{center}
\includegraphics[scale=0.46]{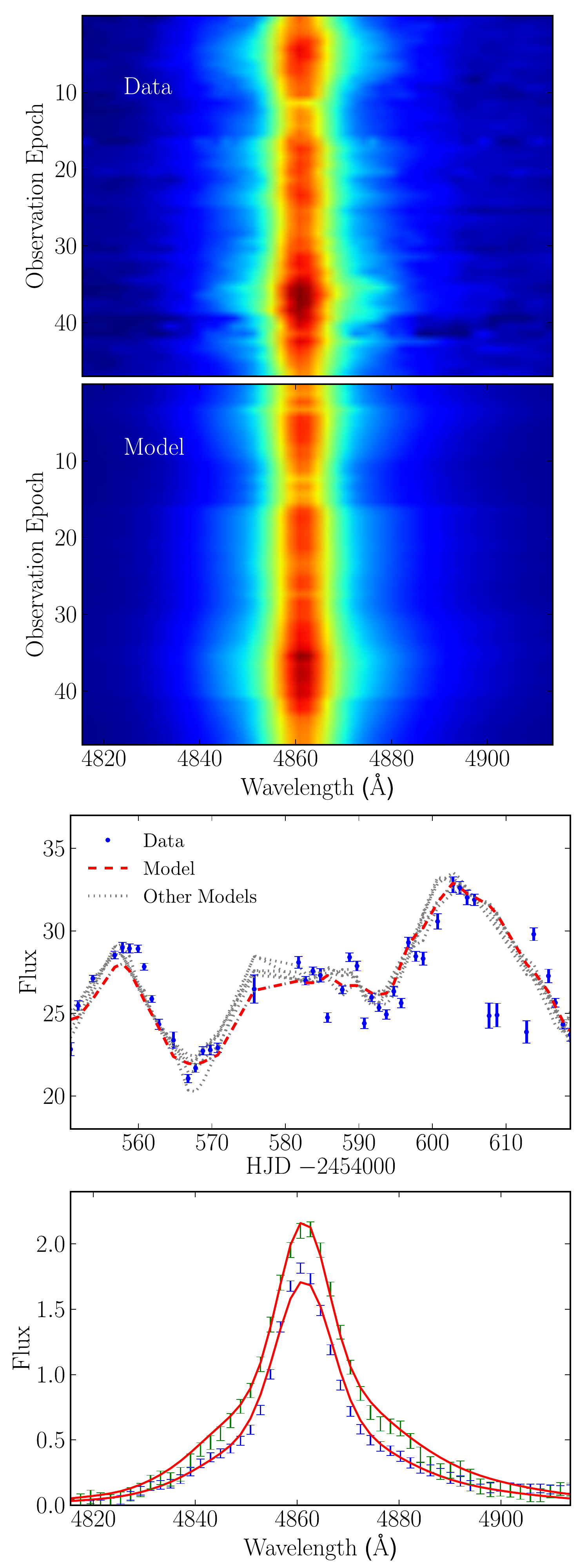}
\caption{Same as Figure~\ref{fig_arp151_rainbow}, but for Mrk 1310.} 
\label{fig_mrk1310_rainbow}
\end{center}
\end{figure}

\begin{figure}
\begin{center}
\includegraphics[scale=0.35]{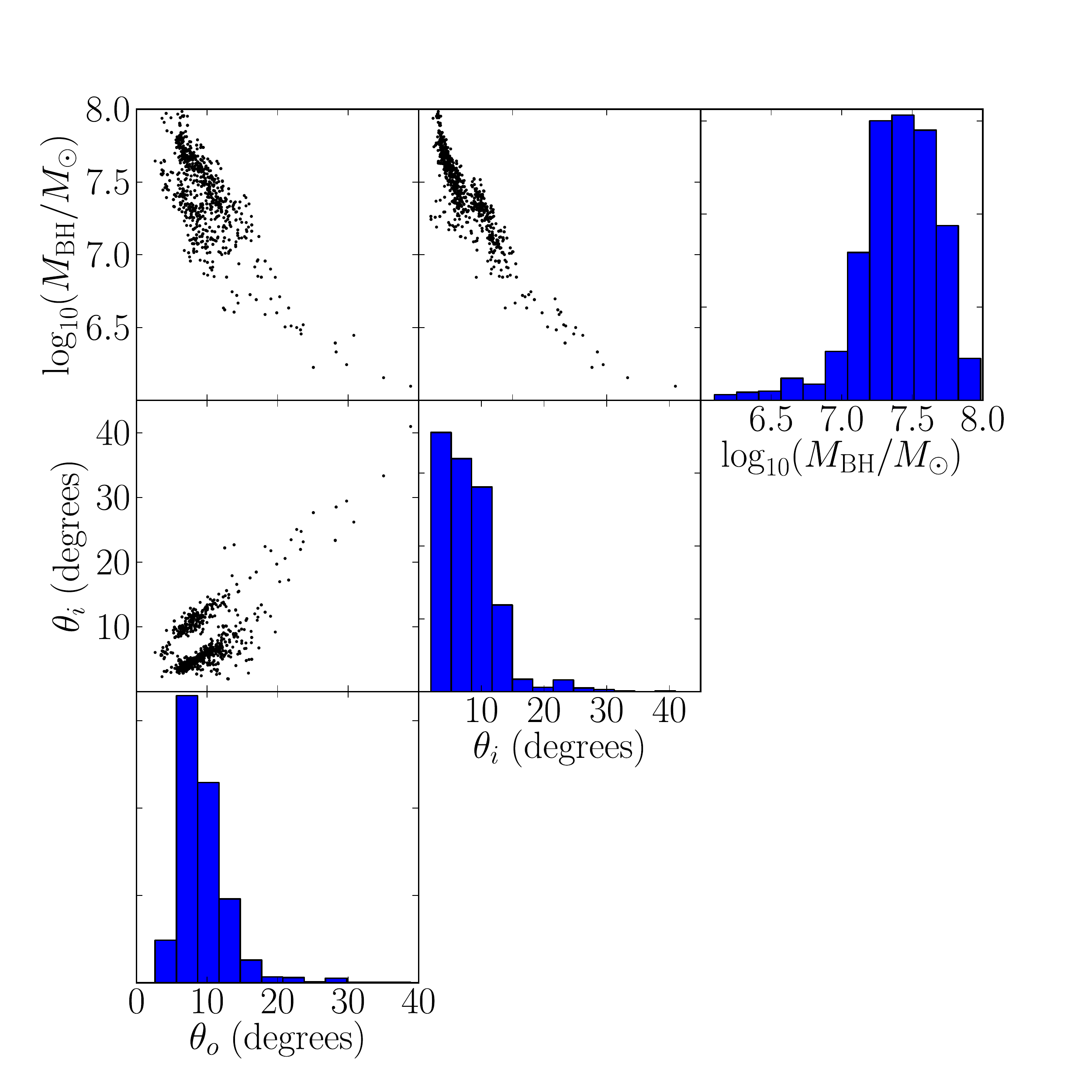}
\caption{Same as Figure~\ref{fig_arp151_cp}, but for Mrk 1310.} 
\label{fig_mrk1310_cp}
\end{center}
\end{figure}

\begin{figure}
\begin{center}
\includegraphics[scale=0.46]{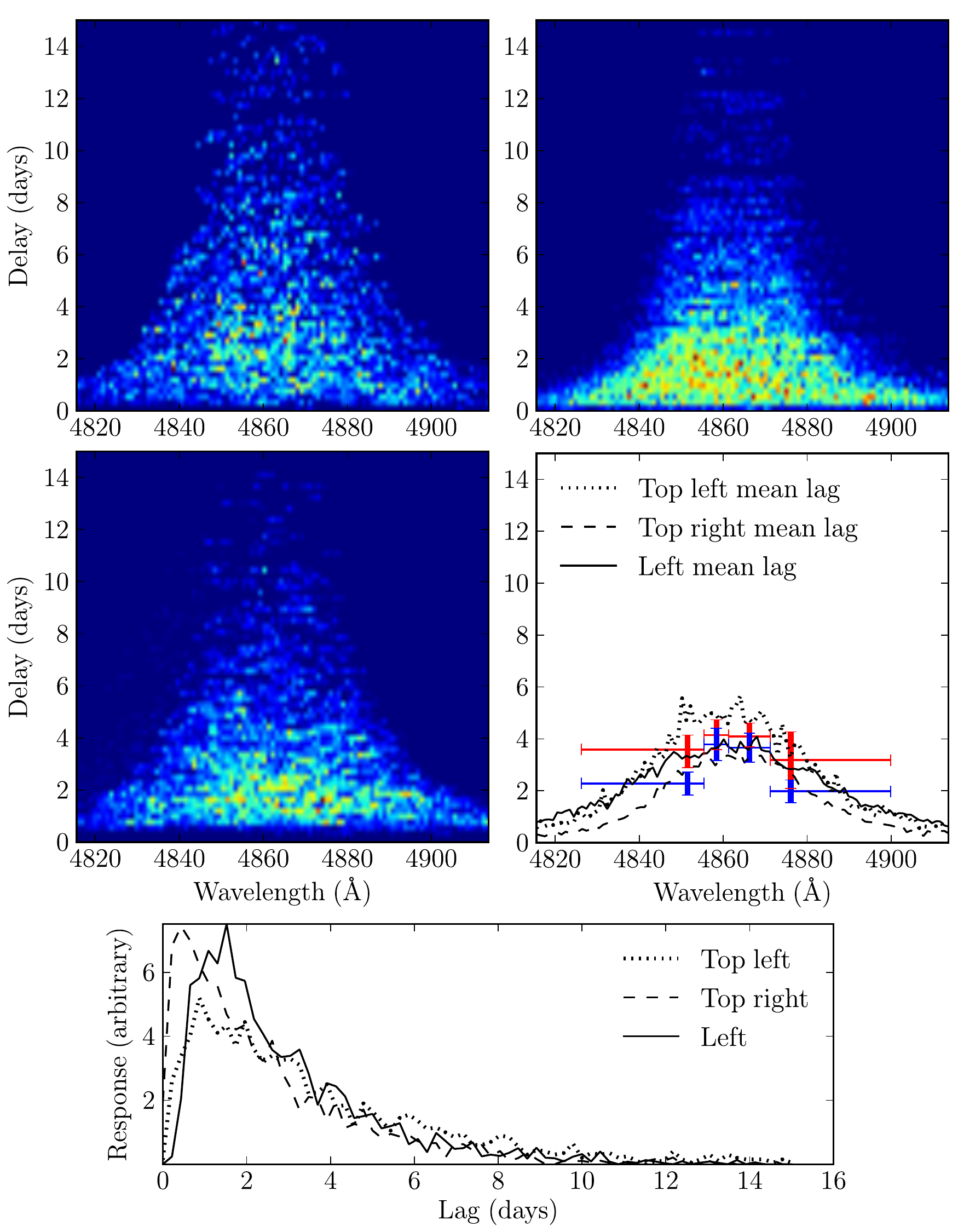}
\caption{Same as Figure~\ref{fig_arp151_tf}, but for Mrk 1310.} 
\label{fig_mrk1310_tf}
\end{center}
\end{figure}

With the narrowest \Hb\ line profile in our sample, the dataset for Mrk 1310
provides fewer constraints on the BLR model due to a smaller number of pixels per spectrum and 
reduced variability compared to Arp 151.  Despite these issues, the model is able to match the overall variability of the emission line profile, 
as well as the detailed line profile shape, as shown in Figure~\ref{fig_mrk1310_rainbow}.  

The geometry of the \Hb\ BLR for Mrk 1310 is constrained to be a thick disk, inclined by 
$\theta_i = 6.6^{+5.0}_{-2.5}$ degrees with respect to the observer, although inclination angles up to 35 degrees
are not completely ruled out.  The radial distribution of \Hb\ emission is constrained to be between 
exponential and Gaussian ($\beta = 0.89^{+0.10}_{-0.10}$), with a mean radius of 
$r_{\rm mean} = 3.13^{+0.42}_{-0.40}$ light days, a minimum radius away from the central source of 
ionizing photons of $r_{\rm min} = 0.12^{+0.19}_{-0.08}$ light days, and a radial dispersion or width
of $\sigma_{r} = 2.59^{+0.42}_{-0.35}$ light days.  The mean time lag of $\tau_{\rm mean} = 2.96^{+0.42}_{-0.35}$ 
days is very similar to the mean radius and median time lag of $\tau_{\rm median} = 2.26^{+0.35}_{-0.31}$, 
and agrees to within the uncertainties with the cross-correlation lag 
of $\tau_{\rm cent} = 3.66^{+0.59}_{-0.61}$ days measured by \citet{bentz09}.  The opening angle of the 
disk is inferred to be $\theta_o = 8.6^{+3.5}_{-2.1}$ degrees, although opening angles up to 35 degrees 
are not completely ruled out.  
There is no preference for \Hb\ emission from the outer faces of the BLR disk ($\gamma = 2.97^{+1.38}_{-1.43}$),
for emission from the far or near side of the BLR ($\kappa = -0.04^{+0.38}_{-0.35}$)
or for the transparency of the BLR midplane ($\xi = 0.40^{+0.38}_{-0.29}$).
An illustration of the BLR geometry for Mrk 1310 is shown in Figure~\ref{fig_geo}
for one sample from the posterior PDF.

The dynamics of the BLR for Mrk 1310 are unclear.  There is a slight preference for elliptical orbits
($f_{\rm ellip} = 0.56^{+0.34}_{-0.39}$) and placement of the inflowing/outflowing velocity distribution 
closer to the distribution centered on the circular orbit value ($\theta_e = 57.2^{+24.9}_{-41.0}$ degrees), 
but also a preference for the remaining orbits to be outflowing when $\theta_e \to 90$ degrees 
($f_{\rm flow} = 0.65^{+0.24}_{-0.38}$, probability of inflow/outflow is 31\%/69\%).
This shows that radial outflowing orbits are not actually strongly preferred compared to radial inflowing orbits 
and the dynamics of the BLR are not well-constrained in this case.
We also find no preference
for substantial macroturbulent velocities ($\sigma_{\rm turb} = 0.004^{+0.010}_{-0.003}$).

The black hole mass for Mrk 1310 is inferred to be $\log_{10}(M_{\rm BH}/M_\odot) = 7.42^{+0.26}_{-0.27}$.  
The uncertainty in the black hole mass is due in large part to degeneracy with the inclination 
angle and opening angle, as shown in Figure~\ref{fig_mrk1310_cp}, since at very small inclination and opening 
angles large changes in black hole mass are needed to maintain the line-of-sight velocity of 
the point particles for even small changes in inclination or opening angle.  Comparing our measurement 
of the black hole mass to the virial products calculated from cross-correlation time lags from \citet{bentz09} 
and line widths from \citet{park12a}, we measure the $f$ factors for Mrk 1310 to be 
$\log_{10}(f_\sigma) = 1.63^{+0.26}_{-0.27}$ and $\log_{10}(f_{\rm FWHM}) = 0.79^{+0.26}_{-0.27}$ (see Section~\ref{sect_arp151}).

The velocity-resolved transfer functions for Mrk 1310, drawn randomly from the posterior PDF, 
show very similar structure as illustrated in Figure~\ref{fig_mrk1310_tf}, 
despite ambiguity in the dynamics of the BLR.  The mean velocity-resolved transfer 
functions and the velocity-integrated transfer functions also show very similar features, and
agree to within the uncertainties with the cross-correlation velocity-resolved lag measurements from \citet{bentz09}.

\subsubsection{NGC 5548}

\begin{figure}
\begin{center}
\includegraphics[scale=0.5]{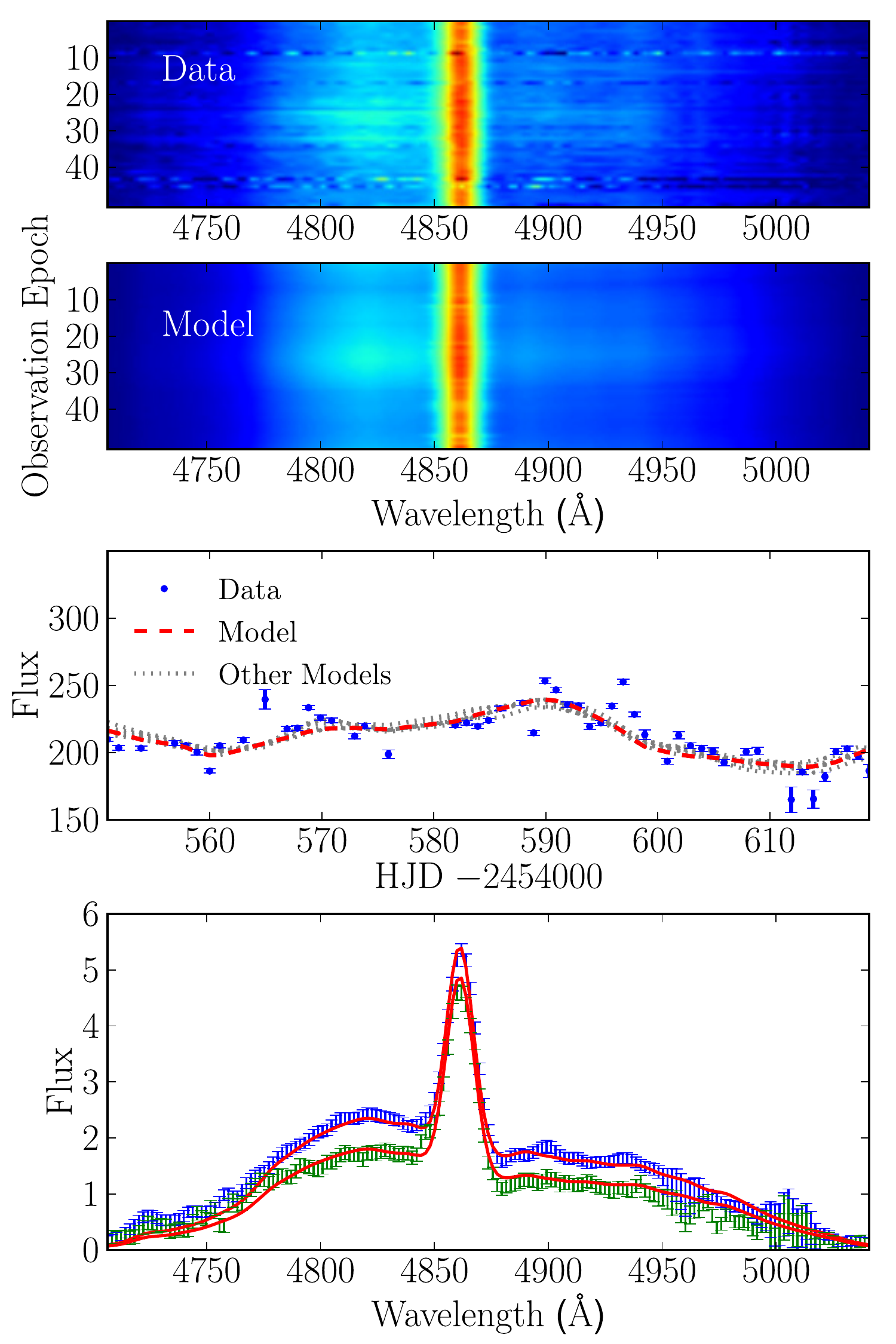}
\caption{Same as Figure~\ref{fig_arp151_rainbow}, but for NGC 5548.} 
\label{fig_ngc5548_rainbow}
\end{center}
\end{figure}

\begin{figure}
\begin{center}
\includegraphics[scale=0.35]{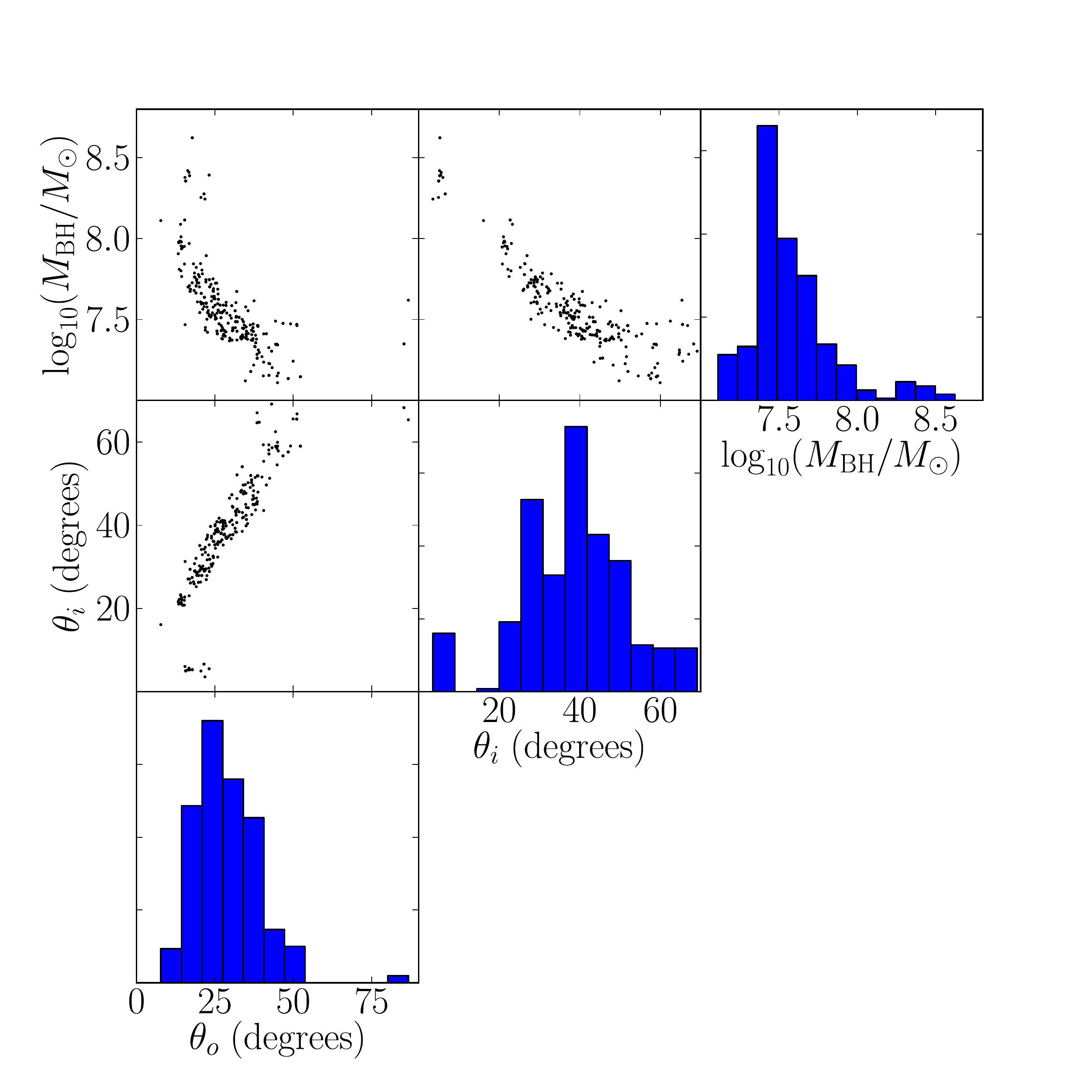}
\caption{Same as Figure~\ref{fig_arp151_cp}, but for NGC 5548.} 
\label{fig_ngc5548_cp}
\end{center}
\end{figure}

\begin{figure}
\begin{center}
\includegraphics[scale=0.46]{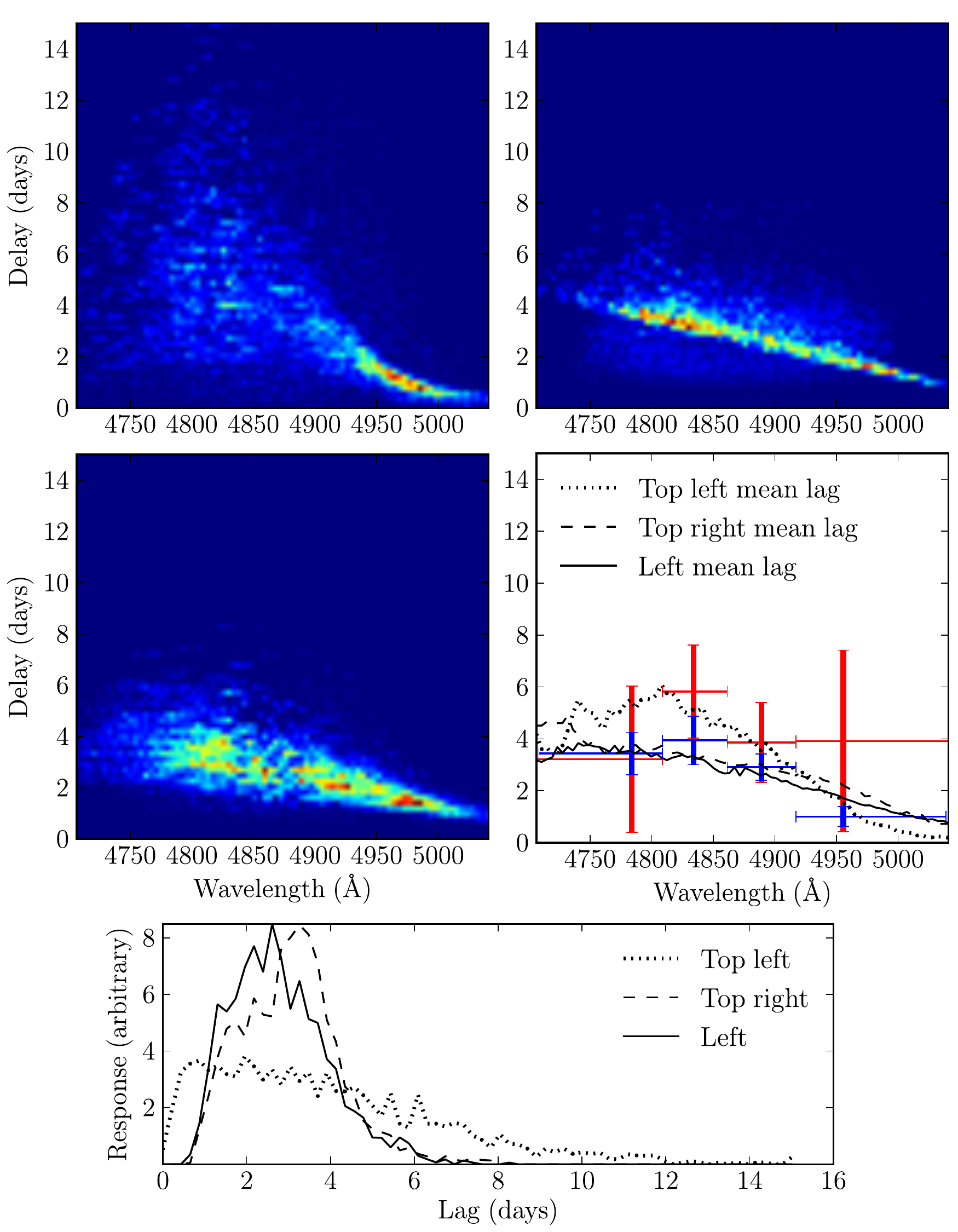}
\caption{Same as Figure~\ref{fig_arp151_tf}, but for NGC 5548.} 
\label{fig_ngc5548_tf}
\end{center}
\end{figure}

While not as variable as Arp 151 over the duration of the LAMP 2008 
campaign, the NGC 5548 \Hb\ line profile
is the widest in the sample, providing an informative dataset with which to 
constrain the BLR model.  The model is able to fit the overall variability of 
the \Hb\ line profile as well as the detailed emission line shape, as shown in 
Figure~\ref{fig_ngc5548_rainbow}.

The geometry of the \Hb\ BLR in NGC 5548 is constrained to be a narrow thick disk, 
with an inclination angle of $\theta_i = 38.8^{+12.1}_{-11.4}$ degrees.  The radial 
distribution of \Hb\ emission is between exponential and Gaussian ($\beta = 0.80^{+0.60}_{-0.31}$), 
with a mean radius of $r_{\rm mean} = 3.31^{+0.66}_{-0.61}$ light days,
a minimum radius from the central ionizing source of $r_{\rm min} = 1.39^{+0.80}_{-1.01}$ light days, 
and a dispersion or width of the BLR of
$\sigma_{r} = 1.50^{+0.73}_{-0.60}$ light days.  The mean lag is very similar to the mean radius, 
with $\tau_{\rm mean} = 3.22^{+0.66}_{-0.54}$ days, and consistent to within
the uncertainties with the cross-correlation lag measurement of $\tau_{\rm cent} = 4.17^{+0.90}_{-1.33}$ 
by \citet{bentz09}.  The median lag is smaller with $\tau_{\rm median} = 2.77^{+0.63}_{-0.42}$ days. The opening angle of the 
disk is inferred to be $\theta_o = 27.4^{+10.6}_{-8.4}$ degrees with opening angles near 90 degrees not completely ruled out and with a slight preference for 
emission equally concentrated throughout the disk ($\gamma = 2.01^{+1.78}_{-0.71}$).  
The \Hb\ emission is also found to preferentially emit from the far side of the BLR 
($\kappa = -0.24^{+0.06}_{-0.13}$) and the midplane of the BLR is found to be not fully transparent
($\xi = 0.34^{+0.11}_{-0.18}$).  An example of the BLR geometry in NGC 5548 is shown in 
Figure~\ref{fig_geo} for a single posterior sample.

The dynamics of the BLR in NGC 5548 are found to be mostly inflow.  The 
fraction of point particles with elliptical orbits is $\sim 10-40\%$ ($f_{\rm ellip} = 0.23^{+0.15}_{-0.15}$), with the
rest of the point particles favoring inflowing orbits ($f_{\rm flow} = 0.25^{+0.21}_{-0.16}$, probability of inflow/outflow is 94\%/6\%).  
Like in the case of Arp 151, the inferred inflowing orbits are mostly bound, with the 
inflow velocity distribution rotated towards the elliptical orbit distribution by $\theta_e = 21.3^{+21.4}_{-14.7}$ 
degrees in the radial and tangential velocities plane.  There is also minimal contribution from additional 
macroturbulent velocities ($\sigma_{\rm turb} = 0.016^{+0.044}_{-0.013}$).

We measure the black hole mass in NGC 5548 to be $\log_{10}(M_{\rm BH}/M_\odot) = 7.51^{+0.23}_{-0.14}$.  
Similar to Arp 151 and Mrk 1310, there are strong correlations between the black hole mass, 
inclination angle, and opening angle that contribute to the uncertainty in black hole mass, as shown in Figure~\ref{fig_ngc5548_cp}.  
By comparing our measurement of the black hole mass to the virial products calculated 
from cross-correlation time lags from \citet{bentz09} and line widths from \citet{park12a}, we 
measure the $f$ factors for NGC 5548 to be $\log_{10}(f_\sigma) = 0.42^{+0.23}_{-0.14}$ and 
$\log_{10}(f_{\rm FWHM}) =  -0.58^{+0.23}_{-0.14}$ (see Section~\ref{sect_arp151}).

The velocity-resolved transfer functions randomly chosen from the posterior show a variety 
of structures consistent with inflow, as shown in Figure~\ref{fig_ngc5548_tf}.  
However, the mean lags for the velocity-resolved transfer functions and the velocity-integrated 
transfer functions are not completely consistent.  Despite this, the velocity-resolved
lag measurements by \citet{bentz09} are consistent to within the uncertainties with our 
mean lag estimates, suggesting that we are able to constrain the general shape of the 
transfer function if not the detailed structure.

\subsubsection{NGC 6814}

\begin{figure}
\begin{center}
\includegraphics[scale=0.46]{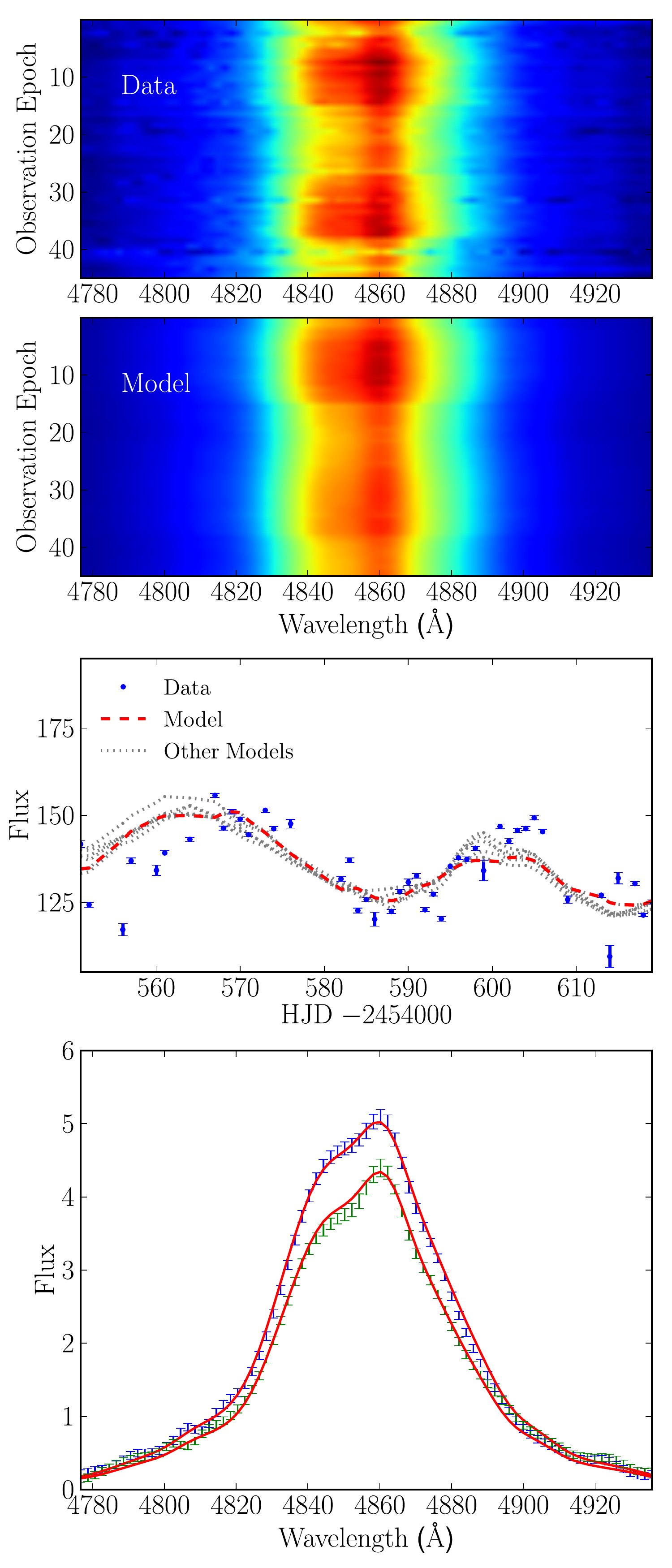}
\caption{Same as Figure~\ref{fig_arp151_rainbow}, but for NGC 6814.} 
\label{fig_ngc6814_rainbow}
\end{center}
\end{figure}

\begin{figure}
\begin{center}
\includegraphics[scale=0.35]{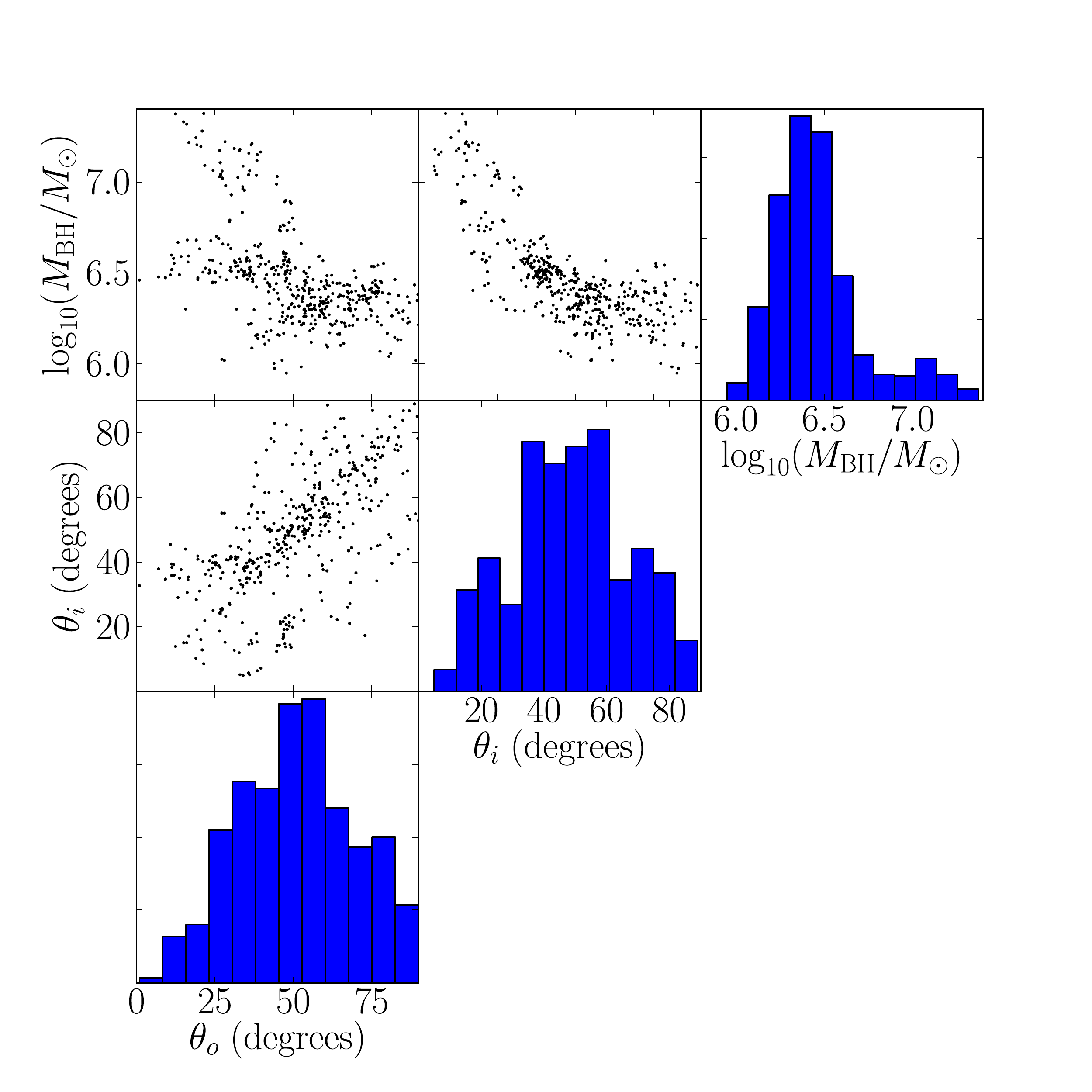}
\caption{Same as Figure~\ref{fig_arp151_cp}, but for NGC 6814.} 
\label{fig_ngc6814_cp}
\end{center}
\end{figure}

\begin{figure}
\begin{center}
\includegraphics[scale=0.46]{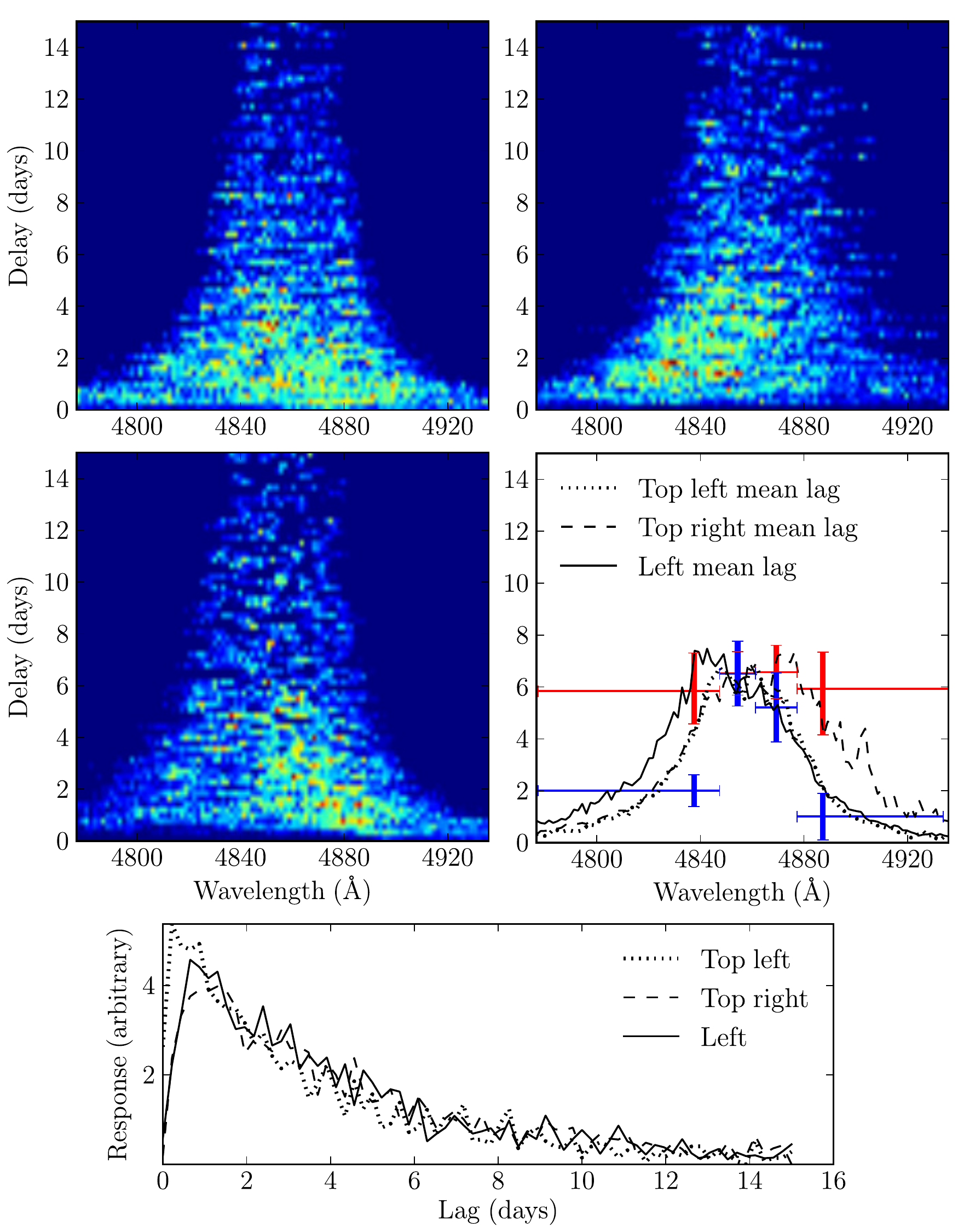}
\caption{Same as Figure~\ref{fig_arp151_tf}, but for NGC 6814.} 
\label{fig_ngc6814_tf}
\end{center}
\end{figure}

While the model is able to capture the detailed line profile shape for NGC 6814, 
it has more difficulty matching the overall variability of the \Hb\ emission, 
as illustrated in Figure~\ref{fig_ngc6814_rainbow}. 
The integrated \Hb\ light curves show some discrepancy, especially at early times, and the 
second bright peak in the spectra is not as strong in the model.

For this object, the BLR as traced by \Hb\ emission is constrained to be a wide thick disk, 
inclined by $\theta_i = 49.4^{+20.4}_{-22.2}$ degrees with respect to the line of sight, where 
inclination angles approaching 90 degrees are not ruled out.  
The radial distribution of \Hb\ emission is close to exponential ($\beta =  1.07^{+0.08}_{-0.09}$), with a mean radius of 
$r_{\rm mean} = 3.76^{+1.15}_{-0.77}$ light days, a minimum radius from the 
central ionizing source of $r_{\rm min} = 0.15^{+0.19}_{-0.11}$ light days, and a dispersion 
or width of the BLR of $\sigma_{r} = 3.75^{+1.05}_{-0.69}$ light days.  The
mean radius is close to the mean time lag of $\tau_{\rm mean} = 4.43^{+0.72}_{-0.83}$ days, which is 
marginally consistent with the cross-correlation lag of $\tau_{\rm cent} = 6.46^{+0.94}_{-0.96}$
by \citet{bentz09}.  The median lag is considerably shorter, with $\tau_{\rm median} = 2.67^{+0.60}_{-0.61}$.  
The opening angle of the disk is inferred to be $\theta_o = 50.2^{+22.0}_{-18.6}$ 
degrees, and a spherical geometry is not ruled out.  While there is no preference
for concentrated \Hb\ emission from the edges of the disk ($\gamma = 2.91^{+1.37}_{-1.31}$), there is a slight
preference for the disk midplane to be transparent ($\xi = 0.71^{+0.22}_{-0.33}$ ) and a  
 strong preference for concentration of \Hb\ emission from the far side of the BLR 
($\kappa = -0.44^{+0.10}_{-0.05}$), although more emission from the near side is not completely ruled out.
The BLR geometry for NGC 6814 from one posterior sample is illustrated in Figure~\ref{fig_geo}.

The dynamics of the BLR for NGC 6814 are inferred to be a combination of elliptical and 
inflowing orbits.  The fraction of elliptical orbits ranges between $0-70\%$ ($f_{\rm ellip} = 0.32^{+0.17}_{-0.22}$),
with the remainder of the orbits mostly inflowing ($f_{\rm flow} = 0.29^{+0.25}_{-0.19}$, probability for inflow/outflow is 83\%/17\%). 
 For the inflowing/outflowing orbits where the fraction of elliptical orbits
is small, the distribution of inflowing/outflowing velocities is rotated by $\sim 60$ degrees 
towards the elliptical orbit distribution in the radial and tangential velocity plane ($\theta_e \sim 60$).  
This means that in the majority of
inferred model solutions with low fractions of elliptical orbits, the inflowing orbits are bound and more similar to 
circular orbits in terms of tangential versus radial velocity component magnitudes.  For the full set of 
posterior model solutions, $\theta_e = 47.0^{+16.7}_{-26.5}$.  Finally, there is minimal contribution from 
additional macroturbulent velocities ($\sigma_{\rm turb} = 0.013^{+0.036}_{-0.011}$).

We measure the black hole mass in NGC 6814 to be $\log_{10}(M_{\rm BH}/M_\odot) = 6.42^{+0.24}_{-0.18}$.  
The correlations of inclination angle and opening angle with black hole mass are not as tight for this object,
adding less uncertainty to the inference of black hole mass, as shown in Figure~\ref{fig_ngc6814_cp}.
By comparing our measurement of the black hole mass to the virial products calculated from 
cross-correlation time lags from \citet{bentz09} and line widths from \citet{park12a}, we measure 
the $f$ factors for NGC 6814 to be $\log_{10}(f_\sigma) =-0.14^{+0.24}_{-0.18}$ and 
$\log_{10}(f_{\rm FWHM}) =  -0.68^{+0.24}_{-0.18}$ (see Section~\ref{sect_arp151}).

The velocity-resolved transfer functions drawn randomly from the
posterior show similar overall structure, as shown in
Figure~\ref{fig_ngc6814_tf}, although an excess of response in the
blue wing, red wing, or center of the line changes between samples.
The line wings also generally have shorter lags than suggested
by the velocity-resolved lag measurements by \citet{bentz09}.  
As for Arp 151, this discrepancy is due to the method of
measuring the time lag.  Again, we confirm this by creating
velocity-resolved light curves using the inferred models of the BLR
for NGC 6814 and comparing the CCF time lag measured from these model
light curves to the CCF time lags from \citet{bentz09}.  In this case,
the comparison is not conclusive. Owing to the low signal-to-noise
ratio of the data in the wings of the line, the cross-correlation
results are very uncertain, and depend significantly upon the details
of the CCF calculation, such as the interval over which the CCF is
calculated.  Despite this, the velocity-integrated transfer functions
are consistent, suggesting that the general shape of the velocity
resolved transfer function is well constrained.

\subsubsection{SBS 1116+583A}

\begin{figure}
\begin{center}
\includegraphics[scale=0.46]{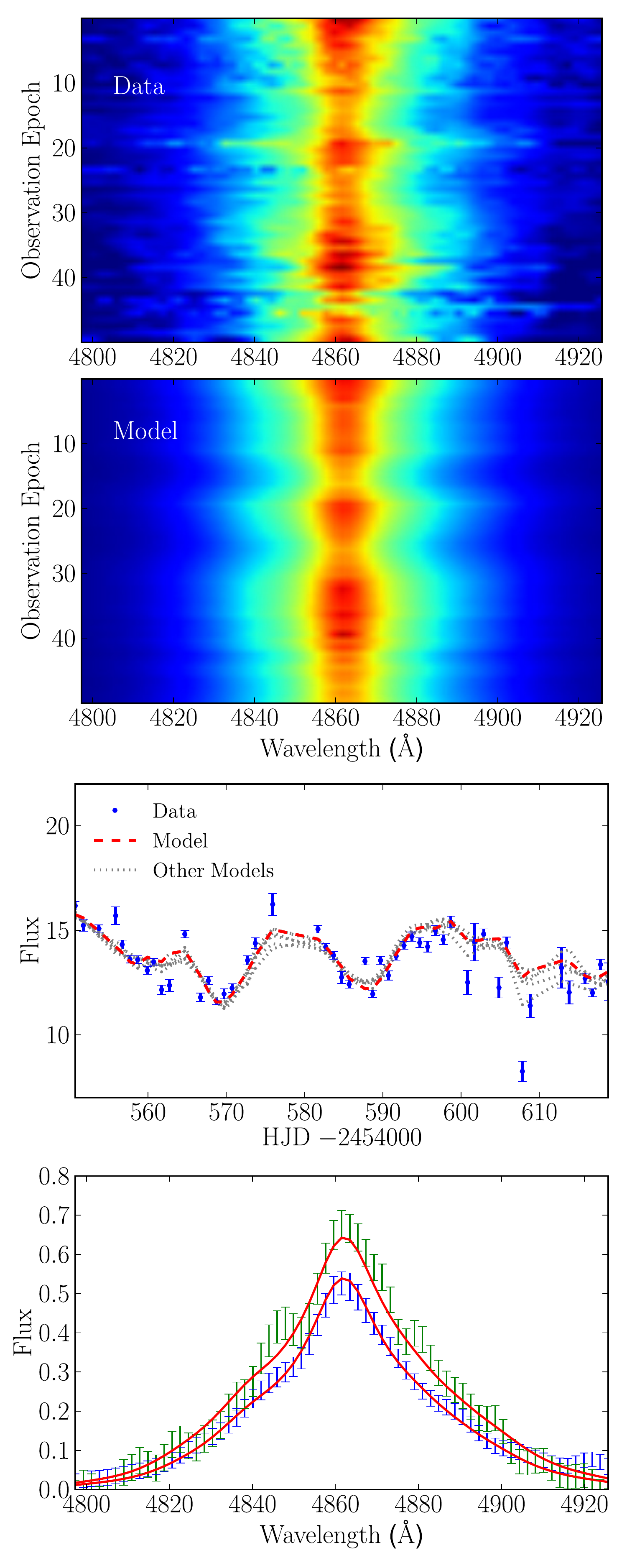}
\caption{Same as Figure~\ref{fig_arp151_rainbow}, but for SBS 1116+583A.} 
\label{fig_sbs1116_rainbow}
\end{center}
\end{figure}

\begin{figure}
\begin{center}
\includegraphics[scale=0.35]{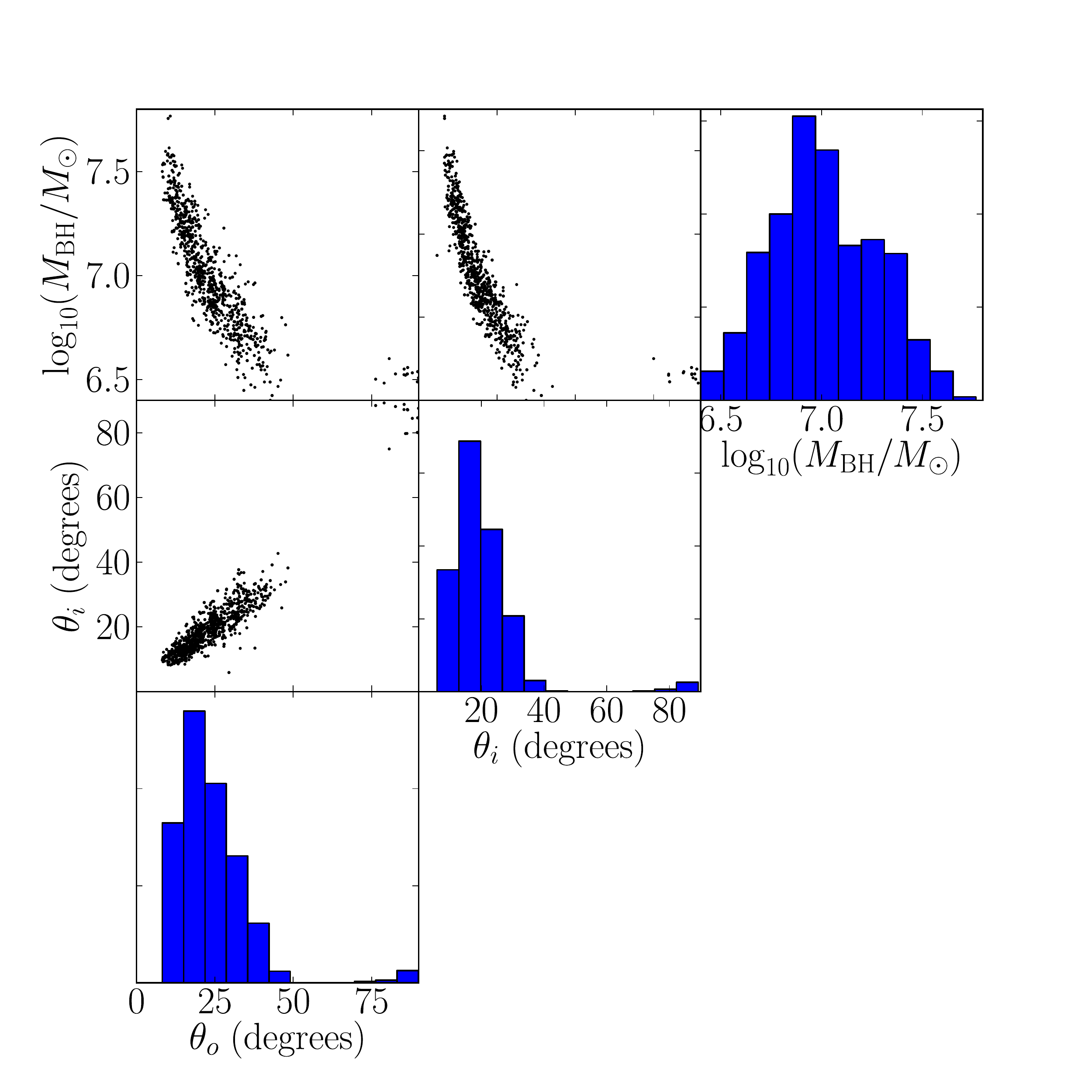}
\caption{Same as Figure~\ref{fig_arp151_cp}, but for SBS 1116+583A.} 
\label{fig_sbs1116_cp}
\end{center}
\end{figure}

\begin{figure}
\begin{center}
\includegraphics[scale=0.46]{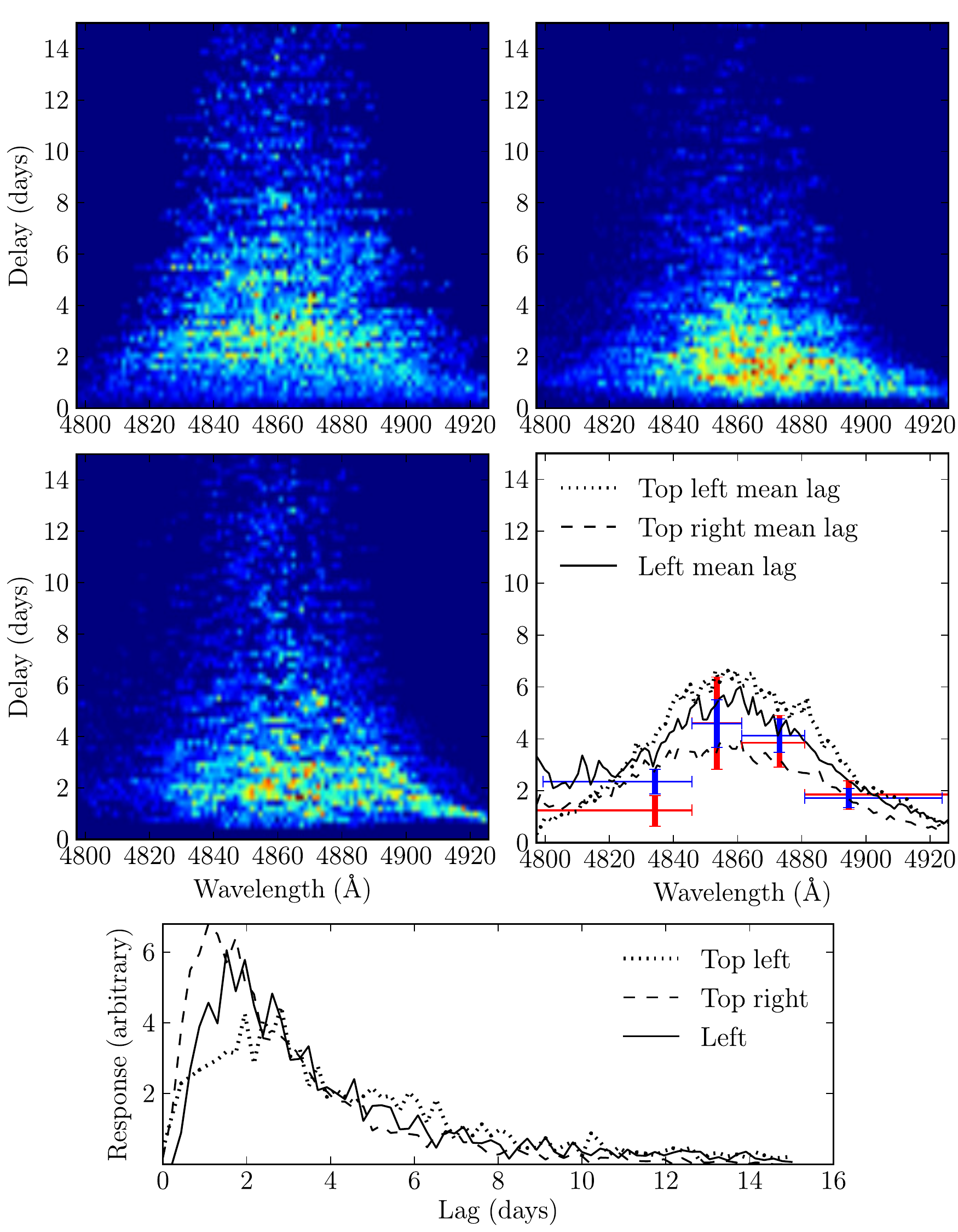}
\caption{Same as Figure~\ref{fig_arp151_tf}, but for SBS 1116+583A.} 
\label{fig_sbs1116_tf}
\end{center}
\end{figure}

The model fits to SBS 1116+583A capture the overall variability of the data and successfully 
match the \Hb\ line profile shape, as shown in Figure~\ref{fig_sbs1116_rainbow}.
We infer the geometry for the BLR in this object to be a wide, thick disk inclined by $\theta_i = 18.2^{+8.4}_{-5.9}$ 
degrees with respect to the line of sight, although inclination angles
approaching 90 degrees are not ruled out.  The radial distribution of \Hb\ emission 
is constrained to be close to exponential ($\beta = 1.00^{+0.27}_{-0.21}$).
The mean radius is $r_{\rm mean} = 4.07^{+0.79}_{-0.65}$ light days, the minimum radius 
from the central ionizing source is $r_{\rm min} = 0.93^{+0.50}_{-0.49}$ light days, and 
the radial dispersion or width of the BLR is $\sigma_{r} = 3.14^{+0.81}_{-0.66}$ light days.  
The mean radius agrees well with the mean lag of $\tau_{\rm mean} = 3.78^{+0.57}_{-0.52}$ days, which
is marginally consistent to within the uncertainties with the cross-correlation lag of 
$\tau_{\rm cent} = 2.31^{+0.62}_{-0.49}$ days \citep{bentz09}. In this case the cross-correlation lag 
is closer to the median time lag of  $\tau_{\rm median} = 2.71^{+0.40}_{-0.37}$ days.  The opening angle of the disk is inferred
to be $\theta_o = 21.7^{+11.0}_{-7.5}$ degrees, and opening angles approaching 90 degrees, 
corresponding to spherical geometries, are not ruled out.  
The other parameters of
the BLR geometry model are unconstrained, 
including emission from 
the front or back side of the BLR ($\kappa = -0.03^{+0.31}_{-0.34}$),
preferential emission from the faces of the disk ($\gamma = 3.19^{+1.21}_{-1.37}$), and
the transparency of the disk mid-plane ($\xi = 0.61^{+0.28}_{-0.37}$). 

The dynamics of the BLR are inferred to be dominated by elliptical orbits.  The 
elliptical orbit fraction is $f_{\rm ellip} = 0.66^{+0.21}_{-0.27}$.  The remaining
orbits are mostly inflowing ($f_{\rm flow} = 0.31^{+0.31}_{-0.22}$, probability of inflow/outflow is 79\%/21\%).  When the elliptical orbit fraction drops below
$\sim50\%$ then the majority of inflow or outflow solutions have $\theta_e > 50$ 
degrees, so the inflow or outflow velocity distributions are rotated in the radial
and tangential velocity plane towards the elliptical orbit distribution.  This is 
compared to $\theta_e = 49.7^{+28.8}_{-32.1}$ degrees for the full posterior.
This means that even posterior samples with a majority of point particle 
velocities drawn from the inflow or outflow
velocity distributions have mainly elliptical-like orbits.  Finally, the dynamics in SBS 1116+583A 
is inferred to have minimal contribution from macroturbulent velocities
with $\sigma_{\rm turb} = 0.011^{+0.033}_{-0.009}$ in units of the circular velocity.  

We measure the black hole mass in SBS 1116+583A to be $\log_{10}(M_{\rm BH}/M_\odot) = 6.99^{+0.32}_{-0.25}$.  
There is a strong correlation between black hole mass and inclination angle and opening angle,
 as shown in Figure~\ref{fig_sbs1116_cp}.  
Comparing our measurement of the black hole mass to the virial products calculated from 
cross-correlation time lags from \citet{bentz09} and line widths from \citet{park12a}, we 
measure the $f$ factors for SBS 1116+583A to be $\log_{10}(f_\sigma) = 0.96^{+0.32}_{-0.25}$ 
and $\log_{10}(f_{\rm FWHM}) =   0.34^{+0.32}_{-0.25}$ (see Section~\ref{sect_arp151}).

Three velocity-resolved transfer functions drawn randomly from the posterior and shown in 
Figure~\ref{fig_sbs1116_tf} show similar detailed structure.  
However, the strength of the prompt emission in the red wing varies
between the velocity-resolved transfer functions, most prominent in the middle left
panel of Figure~\ref{fig_sbs1116_tf} and least prominent in the top left panel.  This is
due to the variation in $f_{\rm ellip}$ and a preference for the remaining orbits to be inflowing.
The velocity-integrated transfer functions also show consistent results, although the 
peakiness of the transfer function at lags of $\sim 1$ day varies.

\subsection{Overview of Modeling Results}

\begin{figure}
\begin{center}
\includegraphics[scale=0.43]{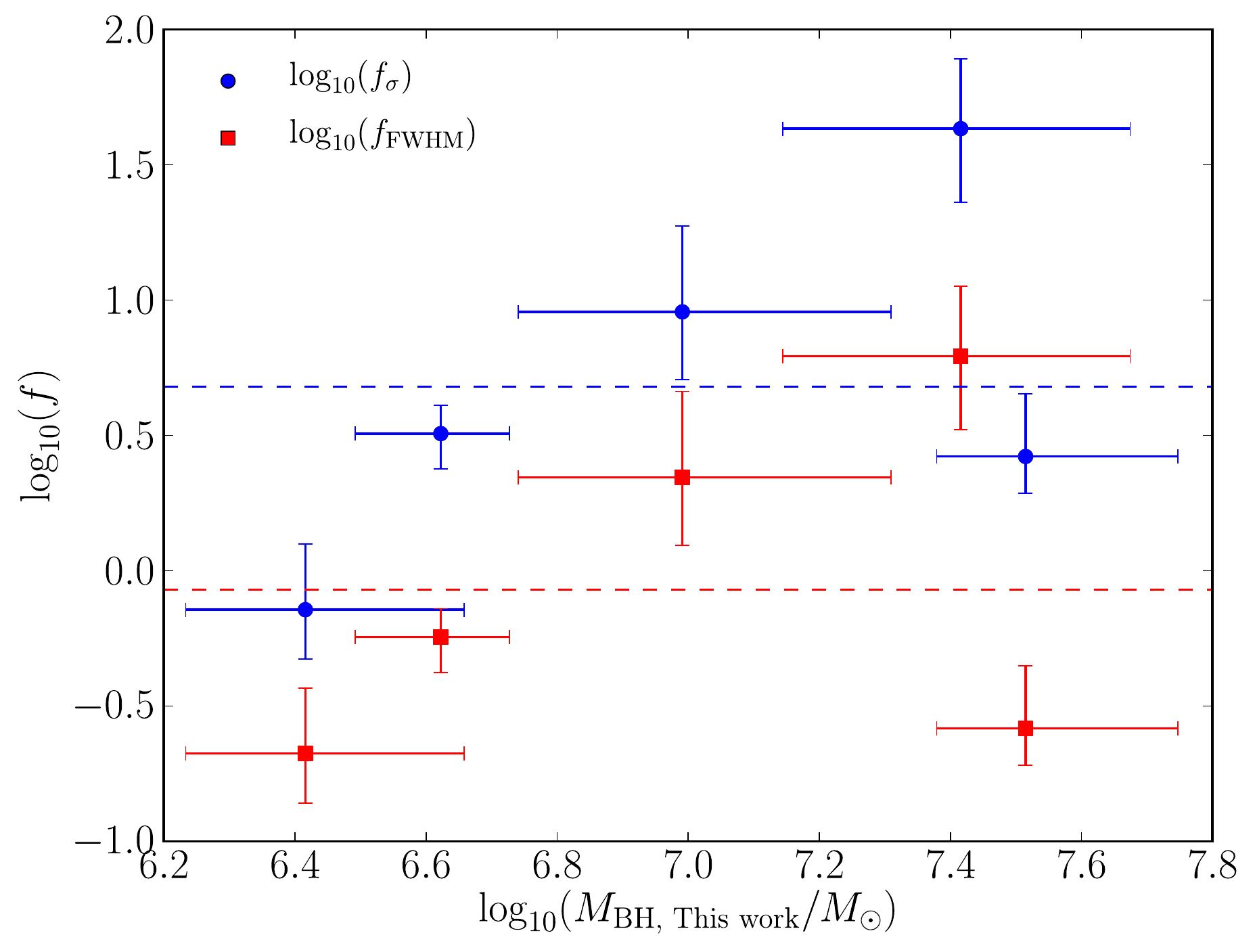}
\caption{Individual $f$ factors versus black hole mass as inferred by direct modeling.  The blue circles show the values of $\log_{10}(f_\sigma)$, while the red squares show the values
of $\log_{10}(f_{\rm FWHM})$.  The values of $\langle \log_{10}(f_\sigma) \rangle$ and $\langle \log_{10}(f_{\rm FWHM}) \rangle$ for the sample are shown by the top blue and bottom red dashed lines, respectively. } 
\label{fig_fmbh}
\end{center}
\end{figure}

\begin{figure}
\begin{center}
\includegraphics[scale=0.43]{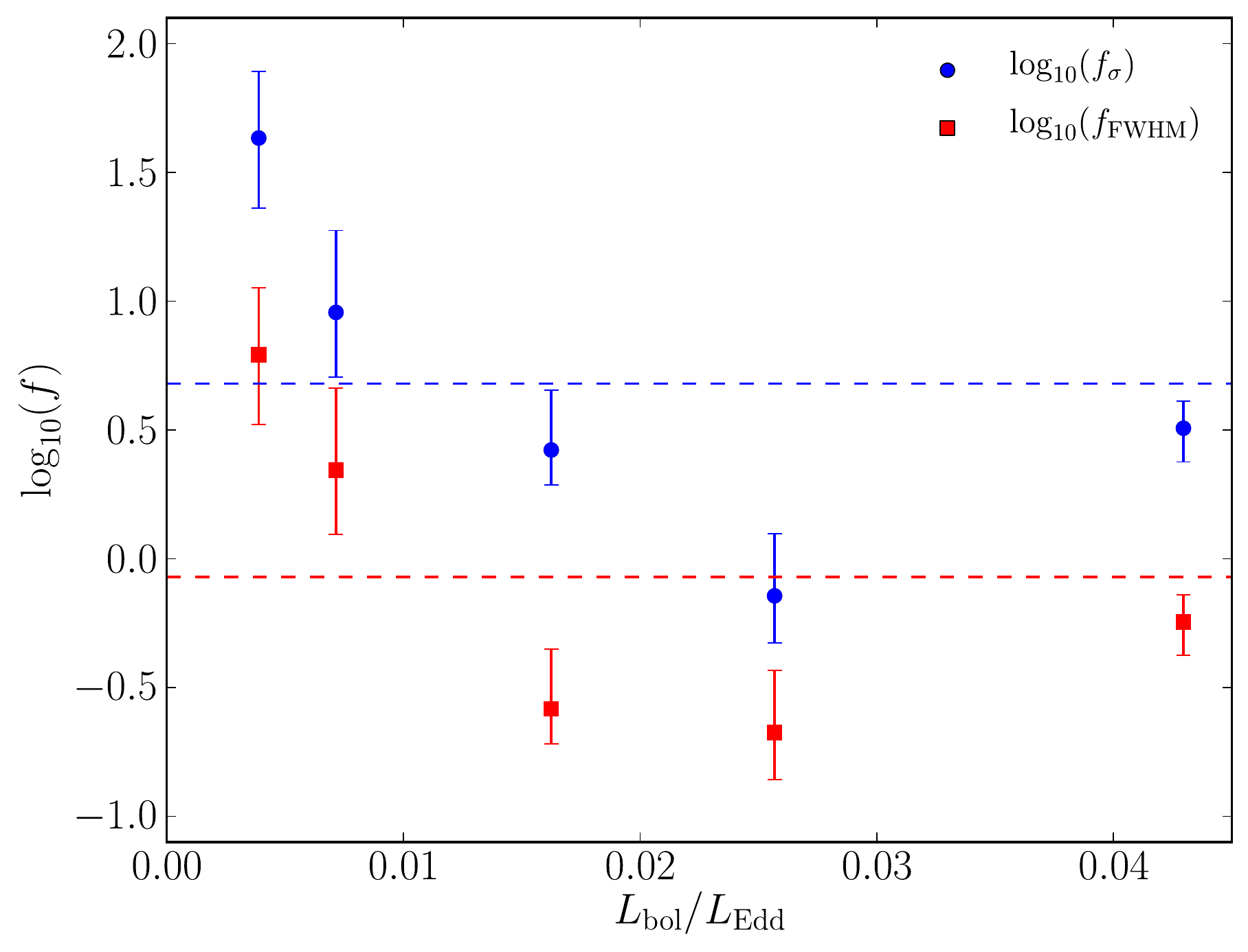}
\includegraphics[scale=0.43]{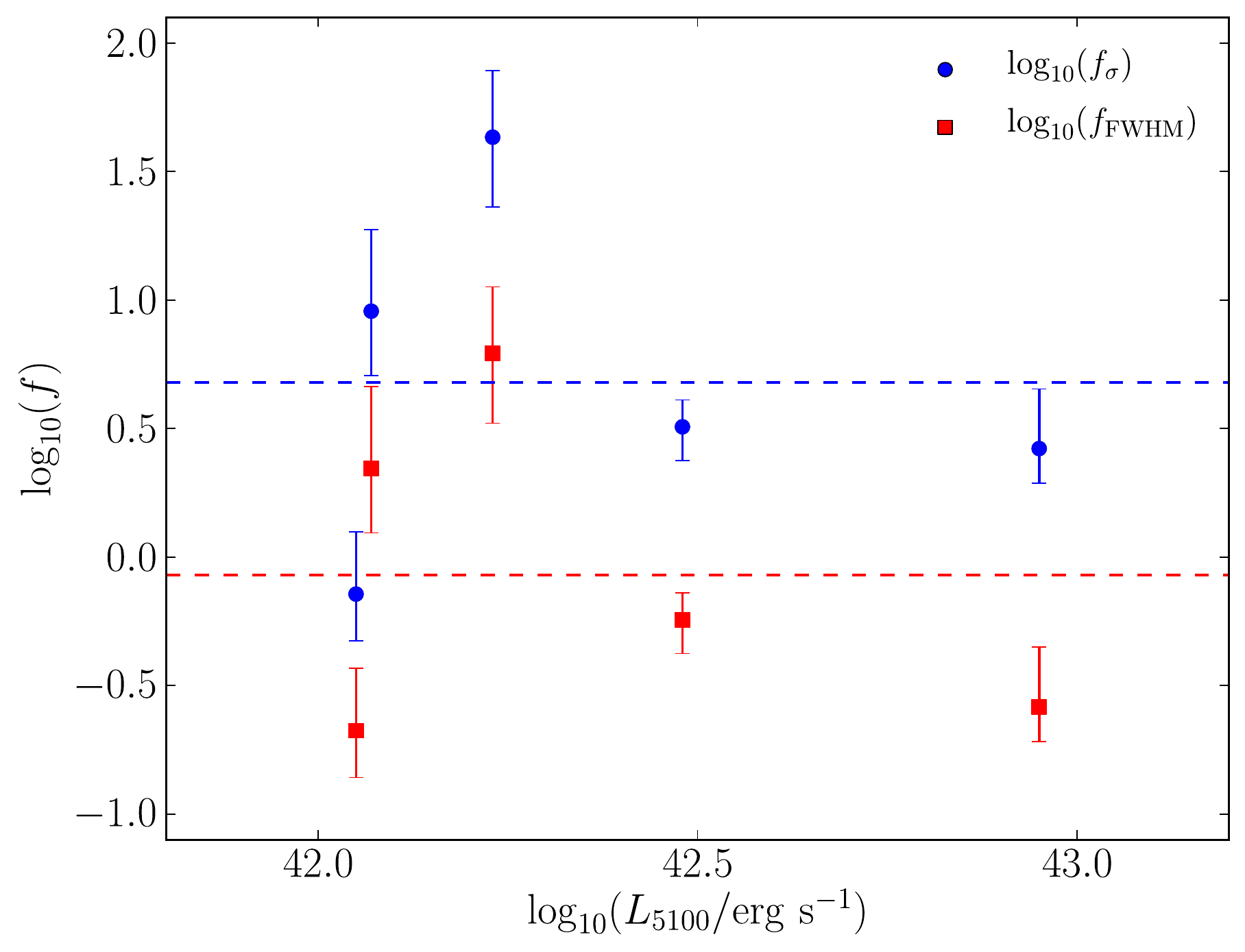}
\caption{ Individual $f$ factors versus Eddington ratio (top panel) and AGN continuum luminosity, $L_{5100}$ (bottom panel). The AGN luminosities at $5100$\AA\ are corrected for host galaxy contamination as described by \citet{bentz13} and the bolometric luminosities are calculated using a bolometric correction factor of nine.   The blue circles show the values of $\log_{10}(f_\sigma)$, while the red squares show the values of $\log_{10}(f_{\rm FWHM})$.  The values of $\langle \log_{10}(f_\sigma) \rangle$ and $\langle \log_{10}(f_{\rm FWHM}) \rangle$ for the sample are shown by the top blue and bottom red dashed lines, respectively.} 
\label{fig_fedd}
\end{center}
\end{figure}

\begin{figure}
\begin{center}
\includegraphics[scale=0.5]{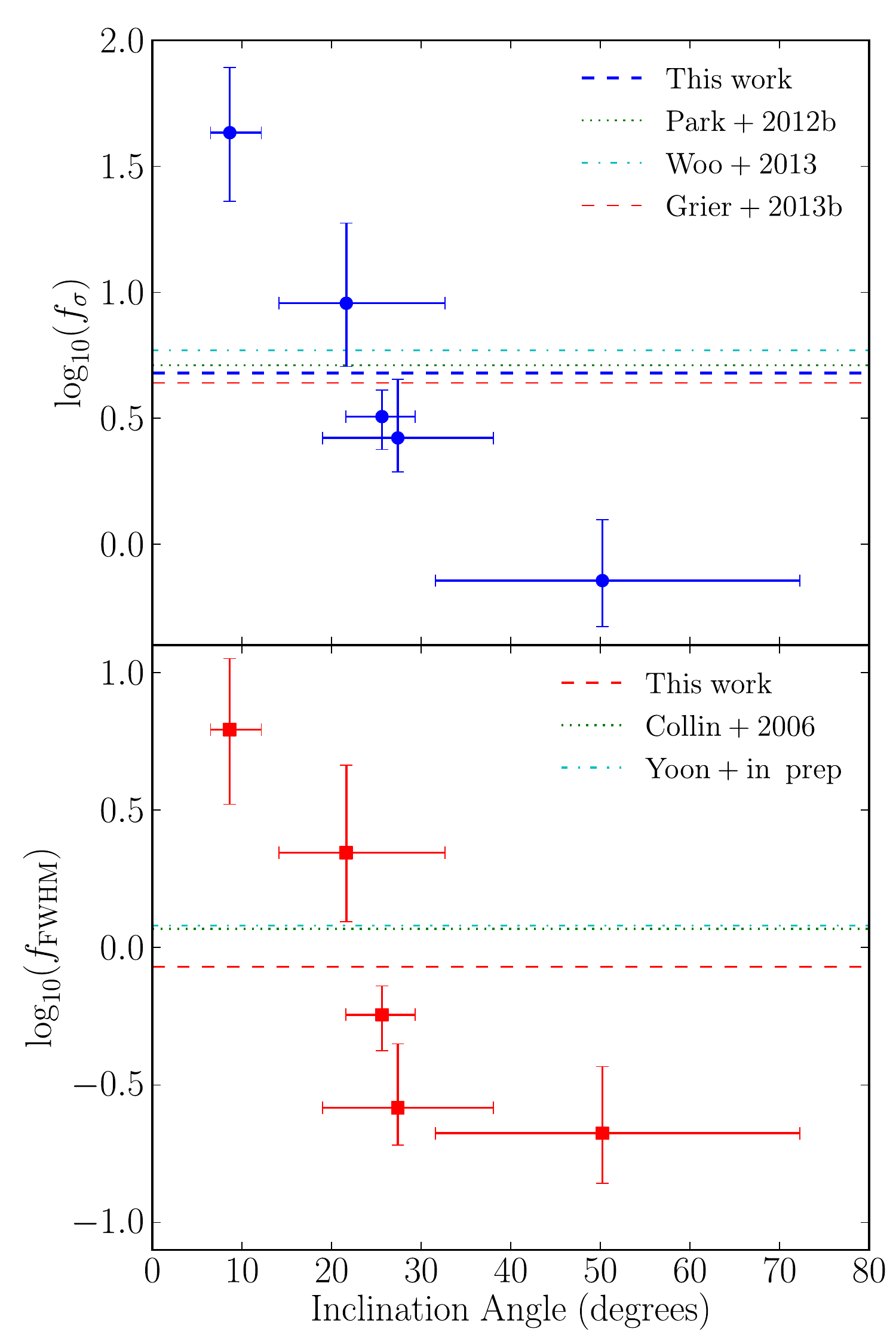}
\caption{Individual $f$ factors versus inclination angle as inferred by direct modeling.  The values of $\langle \log_{10}(f_\sigma) \rangle$ and $\langle \log_{10}(f_{\rm FWHM}) \rangle$ for the sample are shown by the blue dashed line in the top panel and the red dashed line in the bottom panel, respectively.  Also shown as horizontal
dotted, dot-dashed, and dashed lines are the mean $f$ values by \citet{park12a},
\citet{woo13}, and \citet{grier13b} in the top panel, and values by \citet{collin06} and Yoon et al. (in preparation) in the bottom panel.} 
\label{fig_fi}
\end{center}
\end{figure}

\begin{figure}
\begin{center}
\includegraphics[scale=0.5]{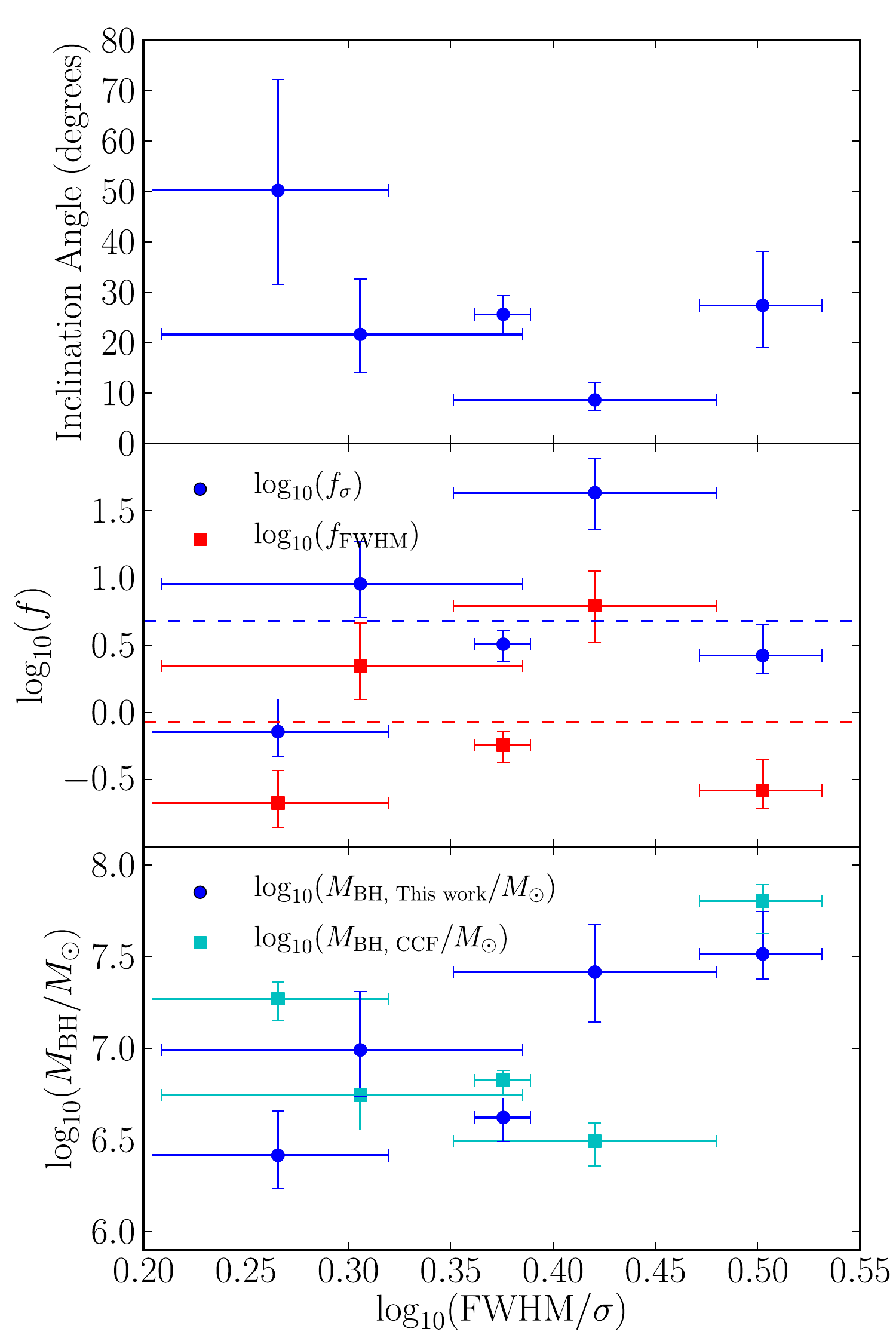}
\caption{Inclination angle, $f$, and black hole mass as a function of the ratio of the FWHM to the line dispersion $\sigma$ of the 
\Hb\ line.  Line width measurements are from \citet{park12a}.  For a Gaussian distribution, $\log_{10}({\rm FWHM}/\sigma) = 0.37$.
 In the middle panel, the values of $\langle \log_{10}(f_\sigma) \rangle$ and $\langle \log_{10}(f_{\rm FWHM}) \rangle$ for the sample are shown by the top blue and bottom red dashed lines, respectively.  In the bottom panel the black hole masses from cross-correlation function analysis and assuming $\log_{10}\langle f_\sigma \rangle = 0.71$ are plotted
 as light blue squares for comparison.} 
\label{fig_f_fwhmsigma}
\end{center}
\end{figure}

\begin{figure}
\begin{center}
\includegraphics[scale=0.43]{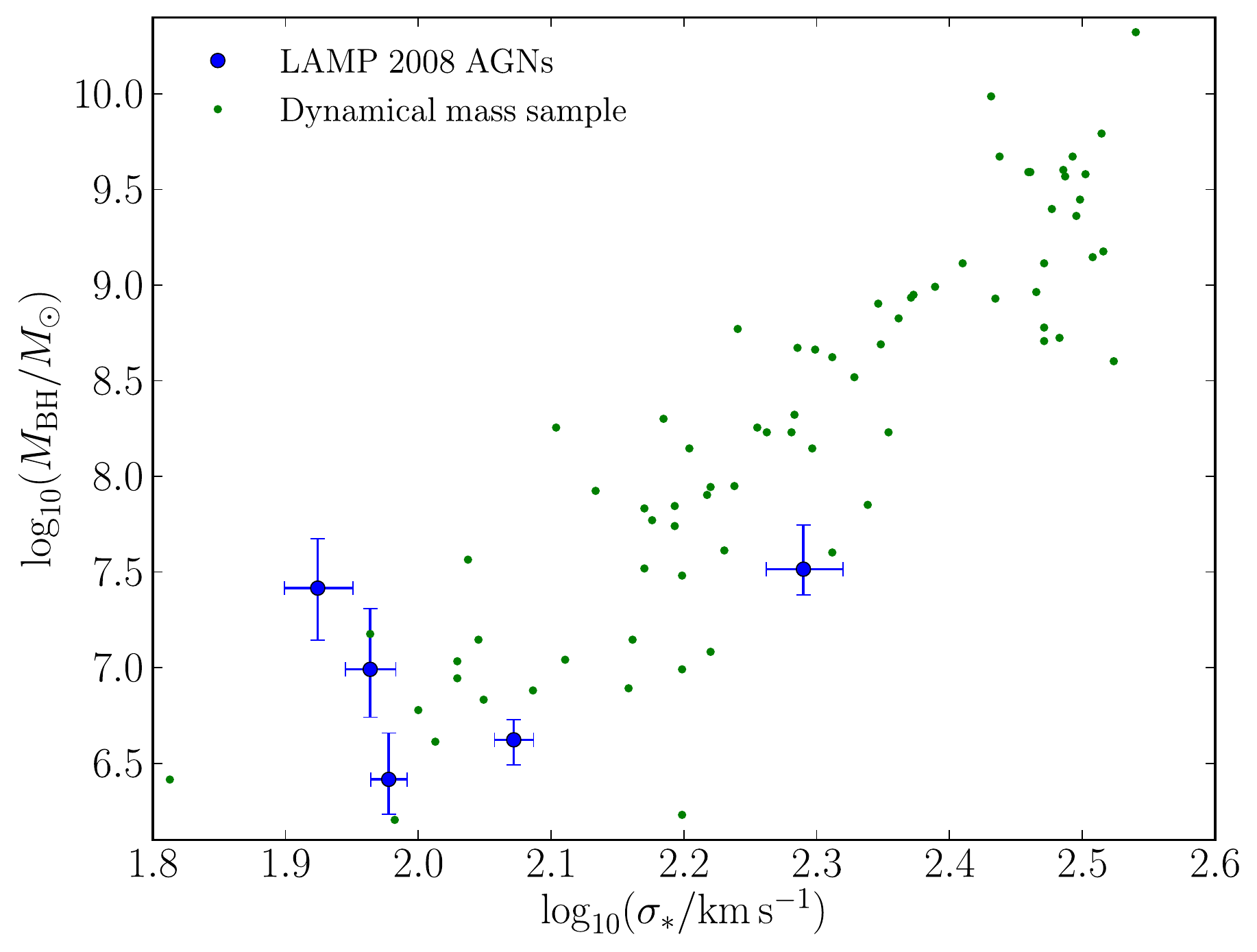}
\caption{
Black hole masses from our direct modeling approach and their host-galaxy stellar velocity dispersions, compared to the
  $M_{\rm BH} - \sigma_*$  relationship for black holes with masses from spatially-resolved stellar and gas dynamical modeling 
  \citep[as compiled by][]{woo13}.
  Measurements of the host-galaxy velocity dispersion for our sample of five LAMP 2008 objects are by \citet{woo10}.  
Our sample is shown by the large blue circles with error bars.  The dynamical mass sample is shown by the small green points. }
\label{fig_msigma}
\end{center}
\end{figure}

\begin{figure}
\begin{center}
\includegraphics[scale=0.43]{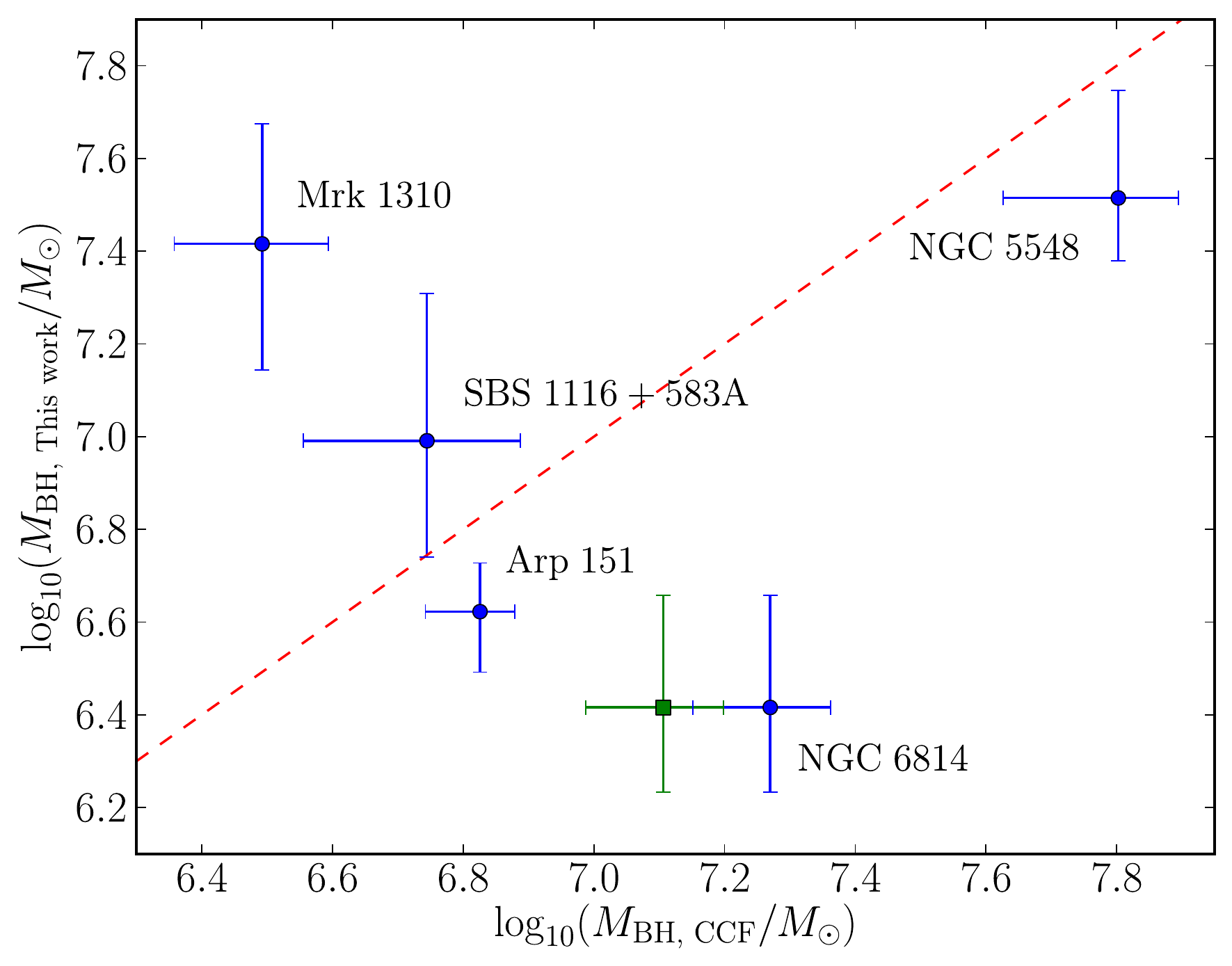}
\caption{Comparison of the black hole mass estimates from our direct modeling approach on the y-axis versus the values measured using cross-correlation function analysis.  
The blue circles denote the sample using cross-correlation 
masses assuming $\log_{10}\langle f_\sigma \rangle = 0.71$, where the points corresponding to individual AGNs have been labeled.
The single green square is for NGC 6814 for the case where the cross-correlation mass has been calculated using the time lag
from our dynamical modeling instead of the time lag from cross-correlation.  The red dashed line shows a slope of unity through the origin.} 
\label{fig_mbh_compare}
\end{center}
\end{figure}

We will now give an overview of the similarities between the inferred BLR model parameters for the five 
objects in our sample.  To begin with, the \Hb\ BLR geometry
is consistent with a thick disk with preferential emission from the far side.  While the minimum 
radius of the BLR from the central ionizing source and the dispersion or width of the 
BLR vary within our sample, the radial distribution shape is generally inferred to be exponential or between 
Gaussian and exponential.    

For the dynamics, we generally infer either elliptical orbits, inflowing orbits, or a combination of the two.  
Both Arp 151 and NGC 5548 show clear signatures of inflow, while SBS 1116+583A shows clear 
signatures of elliptical orbits and NGC 6814 shows evidence for both inflow and elliptical orbits.  In addition, 
both Arp 151 and NGC 5548 prefer bound inflowing orbits, a solution that is closer to
the elliptical orbit solution.  
The absence of strong outflow dynamics in our sample is reassuring, since
reverberation mapping relies on BLR gas dynamics being dominated by the gravitational potential of the black hole,
although this is unsurprising given the low Eddington ratios of the objects in our sample. 

We can also examine whether there are common degeneracies between the model parameters.  The correlations between black hole mass, inclination angle, and opening angle are
typically quite pronounced in our sample (see Figures~\ref{fig_arp151_cp}, \ref{fig_mrk1310_cp}, \ref{fig_ngc5548_cp}, \ref{fig_ngc6814_cp}, and \ref{fig_sbs1116_cp}), and often
the correlation between black hole mass and inclination angle is the strongest.  This degeneracy is very important for BLRs viewed close to face-on, where the uncertainty in black
hole mass becomes larger as the inclination angle approaches zero.  
Smaller opening angles accentuate the degeneracy, leading to strong correlations as for Mrk 1310 (see Figure~\ref{fig_mrk1310_cp}).  
An interesting consequence of these degeneracies is what they predict for correlations of model parameters with individual values of the $f$ factor.  

As shown in Figure~\ref{fig_fmbh}, 
there is no strong correlation between the $f$ factor and black hole mass, as one might 
expect if the BLR geometry and dynamics are somehow correlated with the size of the black hole.
There is also no strong correlation between the $f$ factor and the Eddington ratio, $L_{\rm bol}/L_{\rm Edd}$, or the AGN continuum luminosity at 5100\AA, $L_{5100}$,
as shown in Figure~\ref{fig_fedd}.  The AGN luminosities are corrected for 
host galaxy contamination by \citet{bentz13} and the Eddington ratios are calculated assuming a bolometric correction factor for $L(5100{\rm \AA})$ of nine. 
A correlation between $f$ and the Eddington ratio might be expected if the BLR geometry or dynamics changed substantially
with accretion rate, for example with contributions to the dynamics from radiation pressure.  
Since both $f$ and the Eddington ratio are calculated using the values of $M_{\rm BH}$ inferred from dynamical modeling,
the errors are correlated.  For this reason we also plot $f$ versus  $L_{5100}$, as
shown in the bottom panel of Figure~\ref{fig_fedd}, which does not have correlated errors, although it is not as 
closely related to accretion rate as the Eddington ratio since it has not been normalized by $M_{\rm BH}$.
However, there does appear to be a correlation between the $f$ factor and inclination angle, as illustrated in Figure~\ref{fig_fi}.  
Such a correlation was predicted by \citet{goad12} for a general class of BLR models similar to the ones used in our direct modeling analysis.  
Since the errors in black hole mass and $f$ are the same, 
and since black hole mass correlates so strongly with inclination angle, one might expect to see at least a small trend between the $f$ factor and inclination angle based only on 
correlated errors.  
Direct modeling on a larger sample of AGNs will allow us to quantify the contribution of correlated errors
to the strength of the correlation between inclination angle and $f$.

On a related note, it has been suggested that the ratio of the FWHM to the line dispersion of broad emission lines is 
related to the inclination angle of the BLR to our line of sight \citep{collin06, goad12}.  
We use the FWHM and line dispersion measurements for the objects in our sample from \citet{park12a}
to investigate the possibility of such trends, as shown in Figure~\ref{fig_f_fwhmsigma}.
We find no strong correlation between $\log_{10}({\rm FWHM}/\sigma)$ and the inclination angle or the $f$
factors for individual AGNs, but we do find a tentative correlation between $\log_{10}({\rm FWHM}/\sigma)$
and black hole mass.  The trend of $\log_{10}({\rm FWHM}/\sigma)$ with black hole mass is not
seen for the virial product.
A larger sample of AGNs with direct modeling analysis could clarify the strength of these correlations.
 
There are few independently measured quantities to compare with our direct modeling results.  One of these is measurements of the time lag from cross-correlation function analysis, 
where we find good agreement within the uncertainties.  Recently, \citet{li13} used our direct modeling formalism to develop an independent code to model the geometry
of the BLR.  Their geometry model includes a Gamma distribution for the radial profile of gas, as well as an opening angle and inclination angle.  
In addition, their model includes non-linear response of the broad emission lines to changes in the continuum light curve.  They measure the mean radius of the BLR for our sample
of five AGNs using their geometry modeling code and obtain results that are mostly consistent with the results presented here.  The one object for which our values of mean radius are 
inconsistent is NGC 6814, for which we measure a smaller value than both the mean lag by 
\citet{bentz09} of $\tau_{\rm cent} = 6.46^{+0.94}_{-0.96}$ days and the mean radius by \citet{li13} of $r_{\rm mean} = 6.9 \pm 0.7$ light days.  
The inconsistency  
between direct modeling results for NGC 6814 for the geometry model of \citet{li13} and the dynamical
model implemented here could be caused by using the integrated line profiles versus the full spectral dataset,
since for the full spectral dataset the model must fit not only the mean time lag but also the response as a function of velocity, placing more stringent constraints on 
the value of the mean radius.
There are also a number
of differences between the geometry model used here and the one used by \citet{li13}, the most important being that we do not include
non-linear response of the emission line flux, while \citet{li13} do not include asymmetry parameters such as $\kappa$, $\gamma$, or $\xi$ in their
model.

We can also compare our independent measurements of the black hole mass to those of quiescent and active galaxies with dynamical mass estimates. 
Using host galaxy velocity dispersion measurements by \citet{woo10}, we overlay our five AGNs onto the dynamical mass sample
from \citet{woo13} on the $M_{\rm BH} - \sigma_*$ relation, as shown in Figure~\ref{fig_msigma}.  
The five objects in our sample are consistent with the distribution of masses and stellar velocity dispersions in the dynamical sample, 
confirming that Seyfert 1 galaxies appear to lie on the same $M_{\rm BH} - \sigma_*$ as Seyfert 2 galaxies with black hole mass
measurements from maser kinematics.  With a larger sample of Seyfert 1 galaxies with direct modeling, we can test whether the 
agreement holds across the entire relation.

Another independently measured quantity is the mean $f$ factor, $\langle f \rangle$, measured by aligning the $M_{\rm BH}-\sigma_*$ relations for
quiescent and active galaxies.  We will denote mean $f$ factors that have been calculated using the dispersion of the RMS emission line profile by $\langle f_\sigma \rangle$
and those that have been calculated using the FWHM of the mean emission line profile by $\langle f_{\rm FWHM} \rangle$.
Values of $\langle f_\sigma \rangle$ from the literature
include $\log_{10}\langle f_\sigma \rangle = 0.74^{+0.12}_{-0.17}$ \citep{onken04}, 
$\log_{10}\langle f_\sigma \rangle = 0.72^{+0.09}_{-0.10}$ \citep{woo10}, 
$\log_{10}\langle f_\sigma \rangle = 0.45 \pm 0.09$ \citep{graham11},
$\log_{10}\langle f_\sigma \rangle = 0.71 \pm 0.11$ \citep{park12b},
$\log_{10}\langle f_\sigma \rangle = 0.77 \pm 0.13$ \citep{woo13}, 
and $\log_{10}\langle f_\sigma \rangle = 0.64^{+0.10}_{-0.12}$ \citep{grier13b}.
These values agree to within the uncertainties except for the value by \citet{graham11}, for which 
the discrepancy is explained by sample selection and choice of the independent variable when 
fitting for $f$.  We choose to adopt the \citet{park12b} value of $\log_{10}\langle f_\sigma \rangle = 0.71$ for calculations of the black hole
mass using the virial product, since it is midway between the two most recent 
values of $\log_{10}\langle f_\sigma \rangle$ by \citet{woo13} and \citet{grier13b} and the difference between either measurement and
the \citet{park12b} value is within the quoted error bars. 

The $f_\sigma$ factors measured individually for the five objects in our sample and listed in Table~\ref{table_f} 
are generally consistent to within the uncertainties with all of these values, although the low value of $f_\sigma$ for 
NGC 6814 is only marginally consistent with the higher $\langle f_\sigma \rangle$ values \citep{onken04, woo10, park12b, woo13, grier13b}.  
Part of the discrepancy for NGC 6814 may be due to the difference in time lags
between the value measured from the cross-correlation function of $\tau_{\rm cent} = 6.46^{+0.94}_{-0.96}$ days \citep{bentz09} 
and the value we infer from direct modeling of $\tau_{\rm mean} = 4.43^{+0.72}_{-0.83}$ days.
Using our measurement of the time lag to calculate the virial mass increases the value of 
$f_\sigma$ by 0.16 dex to $\log_{10}(f_\sigma) = 0.02^{+0.24}_{-0.18}$ for NGC 6814.  
To better illustrate this issue, a comparison of our independent 
measurements of black hole mass to those measured using cross-correlation 
function analysis and assuming $\log_{10}\langle f_\sigma \rangle = 0.71$ \citep{park12b} is shown in Figure~\ref{fig_mbh_compare}. 
NGC 6814 has one of the most discrepant measurements of the black 
hole mass, and the discrepancy is reduced when the cross-correlation mass is 
calculated using the smaller time lag we infer from direct modeling.  
However, since the posterior PDF for the black hole mass in NGC 6814 extends up to values of $\log_{10}(M_{\rm BH}/M_\odot) \sim 7.3$,
this means the posterior PDF for $f_\sigma$ also extends up to values consistent with the higher $\langle f_\sigma \rangle$ values.
While the high posterior median value of $f_\sigma$ for Mrk 1310 is also only marginally consistent with the higher $\langle f_\sigma \rangle$ values, 
the posterior PDF for black hole mass for Mrk 1310 extends down to values below $\log_{10}(M_{\rm BH}/M_\odot) \sim 6.5$
and hence the posterior PDF for $f_\sigma$ also extends down to values consistent with the higher $\langle f_\sigma \rangle$ values.

There are fewer measurements of $f_{\rm FWHM}$ in the literature.  \citet{collin06} find
$\log_{10}\langle f_{\rm FWHM} \rangle = 0.07^{+0.15}_{-0.24}$, in good agreement with 
the more recently calculated value of $\log_{10}\langle f_{\rm FWHM} \rangle = 0.08 \pm 0.12$ 
from Yoon et al. (in preparation).  
While three of the AGNs in our sample have values of $f_{\rm FWHM}$ consistent with the mean value of 
Yoon et al., Mrk 1310 and NGC 6814 have values that are only marginally consistent.

\subsection{The Mean $f$ factor for LAMP 2008}

\begin{figure}
\begin{center}
\includegraphics[scale=0.43]{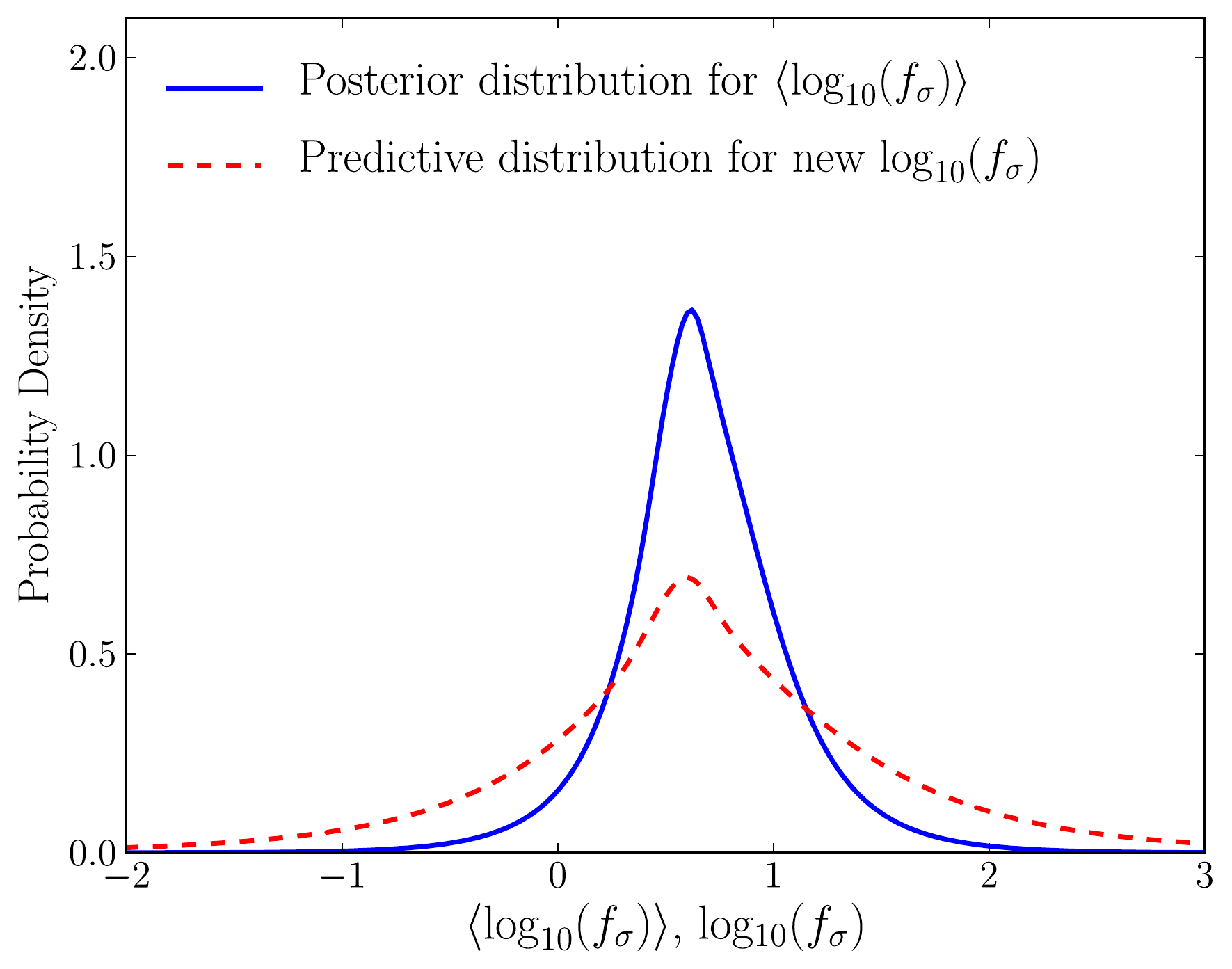}
\includegraphics[scale=0.43]{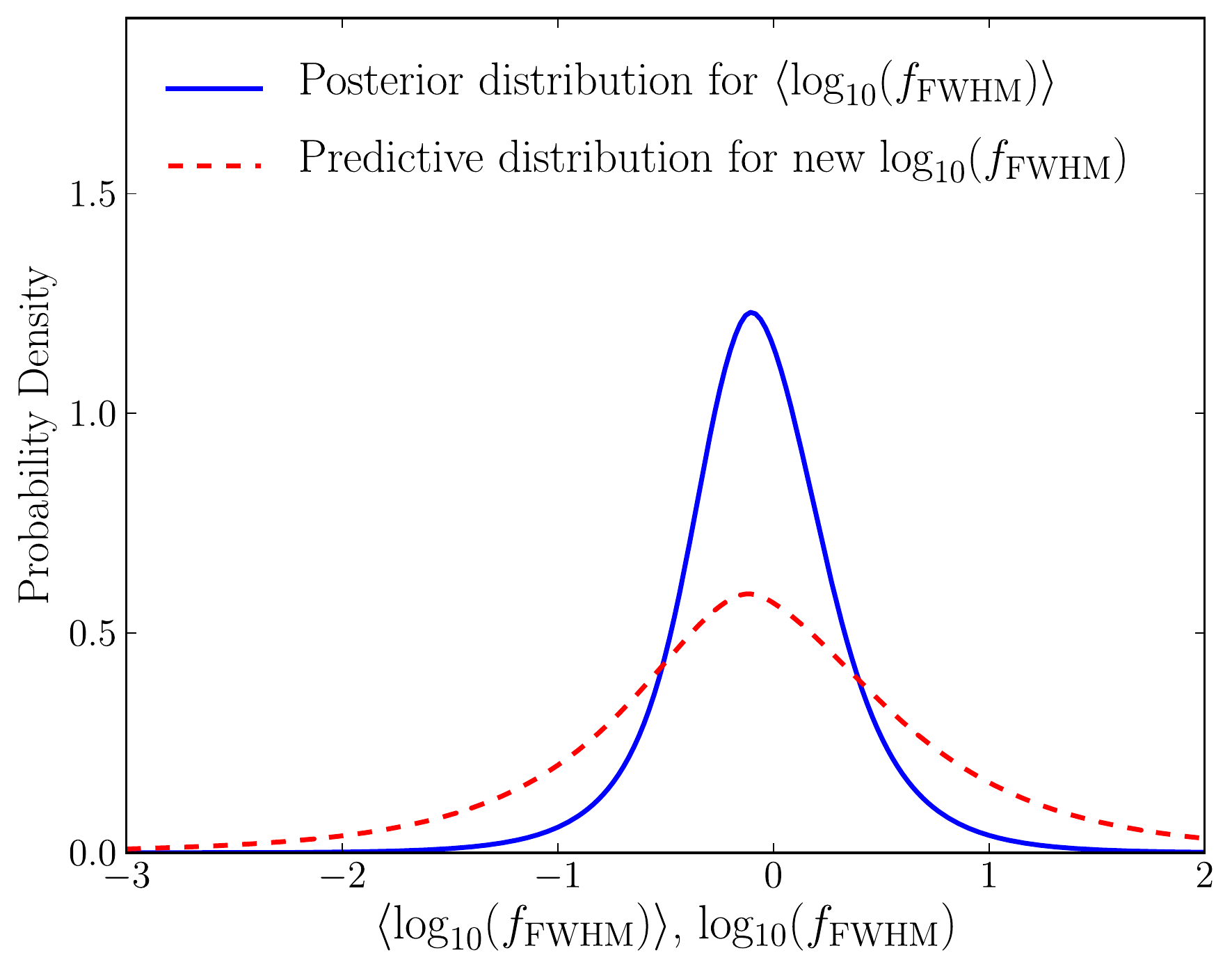}
\caption{Top: The posterior distribution for $\langle \log_{10}(f_\sigma) \rangle$, the mean of the $f_\sigma$ 
factor distribution for our sample of five AGNs, is shown by the solid blue line.  The predictive distribution
for new measurements of $\log_{10}(f_\sigma)$ is shown by the dashed red line.  
Bottom: The posterior distribution for $\langle \log_{10}(f_{\rm FWHM}) \rangle$, the mean of the $f_{\rm FWHM}$ 
factor distribution for our sample of five AGNs, is shown by the solid blue line.  The predictive distribution
for new measurements of $\log_{10}(f_{\rm FWHM})$ is shown by the dashed red line.} 
\label{fig_meanf}
\end{center}
\end{figure}

With five independent black hole mass measurements we can now calculate the mean $f$ factors for our AGN sample,
 called $\langle f_\sigma \rangle$ and $\langle f_{\rm FWHM} \rangle$.
We use the full posterior distributions of $f$ for each AGN to
measure the mean and the dispersion of the distribution of $f$ factors for the whole sample, as described in Appendix~\ref{appendix_f}.
We measure a value for 
$\langle \log_{10}(f_\sigma) \rangle$ of $0.68 \pm 0.40$ and a dispersion for $\log_{10}(f_\sigma)$ of $0.75 \pm 0.40$, while 
we measure a value for 
$\langle \log_{10}f_{\rm FWHM}) \rangle$ of $-0.07 \pm 0.40$ and a dispersion for $\log_{10}(f_{\rm FWHM})$ of $0.77 \pm 0.38$.  
The posterior distributions for $\langle \log_{10}(f_\sigma) \rangle$ and $\langle \log_{10}(f_{\rm FWHM}) \rangle$ and the predictive distributions
for new measurements of $\log_{10}(f_\sigma)$ and $\log_{10}(f_{\rm FWHM})$ are
illustrated in Figure~\ref{fig_meanf}.  The predictive distribution is the 
distribution from which new measurements of $f$ are drawn and is wider than the posterior for the mean value due
to the large scatter in individual $f$ posterior distributions.
Both our values of $\langle f_\sigma \rangle$ and its dispersion are consistent to within the uncertainties with the values for
$\langle f_\sigma \rangle$ measured by aligning the $M_{\rm BH} - \sigma_*$ relation for active galaxies with the relation for galaxies with
dynamical mass estimates \citep[e.g.][]{onken04, woo10, graham11, park12b, woo13, grier13b}.  
Similarly, our values of $\langle f_{\rm FWHM} \rangle$ and its dispersion are consistent to within the uncertainties with the values measured
by \citet{collin06} and Yoon et al. (in preparation).
The mean $f$ factors derived here are meant to illustrate the capabilities of the direct modeling approach and should not be used to normalize
the black hole masses of reverberation mapped AGNs until the direct modeling sample is both larger and more representative of the overall AGN population.

\begin{table*}
 \begin{minipage}{140mm}
  \caption{Inferred posterior median parameter values and central 68\% credible intervals for
direct modeling of five LAMP 2008 AGNs.}
  \label{table_results}
  \begin{tabular}{@{}cccccc}
   \hline
   Geometry Model Parameter & Arp 151  &  Mrk 1310 &  NGC 5548 & NGC 6814  & SBS 1116+583A \\
   \hline
$r_{\rm mean}$ (light days) &   $3.44^{+0.26}_{-0.24}$ &   $3.13^{+0.42}_{-0.40}$ &   $3.31^{+0.66}_{-0.61}$ &   $3.76^{+1.15}_{-0.77}$ &   $4.07^{+0.79}_{-0.65}$ \\ 
$r_{\rm min}$ (light days) &   $0.44^{+0.13}_{-0.20}$ &   $0.12^{+0.19}_{-0.08}$ &   $1.39^{+0.80}_{-1.01}$ &   $0.15^{+0.19}_{-0.11}$ &   $0.93^{+0.50}_{-0.49}$ \\ 
$\sigma_{r}$ (light days) &   $3.72^{+0.45}_{-0.43}$ &   $2.59^{+0.42}_{-0.35}$ &   $1.50^{+0.73}_{-0.60}$ &   $3.75^{+1.05}_{-0.69}$ &   $3.14^{+0.81}_{-0.66}$ \\ 
$\tau_{\rm mean}$  (days)  &   $3.07^{+0.25}_{-0.20}$ &   $2.96^{+0.42}_{-0.35}$ &   $3.22^{+0.66}_{-0.54}$ &   $4.43^{+0.72}_{-0.83}$ &   $3.78^{+0.57}_{-0.52}$ \\ 
$\tau_{\rm median}$  (days)  &   $1.75^{+0.28}_{-0.23}$ &   $2.26^{+0.35}_{-0.31}$ &   $2.77^{+0.63}_{-0.42}$ &   $2.67^{+0.60}_{-0.61}$ &   $2.71^{+0.40}_{-0.37}$ \\ 
$\beta$ &   $1.25^{+0.15}_{-0.16}$ &   $0.89^{+0.10}_{-0.10}$ &   $0.80^{+0.60}_{-0.31}$ &   $1.07^{+0.08}_{-0.09}$ &   $1.00^{+0.27}_{-0.21}$ \\ 
$\theta_o$ (degrees) &   $25.6^{+ 3.7}_{- 4.0}$ &   $ 8.6^{+ 3.5}_{- 2.1}$ &   $27.4^{+10.6}_{- 8.4}$ &   $50.2^{+22.0}_{-18.6}$ &   $21.7^{+11.0}_{- 7.5}$ \\ 
$\theta_i$ (degrees) &   $25.2^{+ 3.3}_{- 3.4}$ &   $ 6.6^{+ 5.0}_{- 2.5}$ &   $38.8^{+12.1}_{-11.4}$ &   $49.4^{+20.4}_{-22.2}$ &   $18.2^{+ 8.4}_{- 5.9}$ \\ 
$\kappa$ &   $-0.36^{+0.08}_{-0.08}$ &   $-0.04^{+0.38}_{-0.35}$ &   $-0.24^{+0.06}_{-0.13}$ &   $-0.44^{+0.10}_{-0.05}$ &   $-0.03^{+0.31}_{-0.34}$ \\ 
$\gamma$ &   $4.27^{+0.54}_{-0.80}$ &   $2.97^{+1.38}_{-1.43}$ &   $2.01^{+1.78}_{-0.71}$ &   $2.91^{+1.37}_{-1.31}$ &   $3.19^{+1.21}_{-1.37}$ \\ 
$\xi$ &   $0.09^{+0.08}_{-0.05}$ &   $0.40^{+0.38}_{-0.29}$ &   $0.34^{+0.11}_{-0.18}$ &   $0.71^{+0.22}_{-0.33}$ &   $0.61^{+0.28}_{-0.37}$ \\ 
   \hline      
   Dynamical Model Parameter & Arp 151  &  Mrk 1310 &  NGC 5548 & NGC 6814  & SBS 1116+583A \\
   \hline
$\log_{10}(M_{\rm BH}/M_\odot)$ &   $6.62^{+0.10}_{-0.13}$ &   $7.42^{+0.26}_{-0.27}$ &   $7.51^{+0.23}_{-0.14}$ &   $6.42^{+0.24}_{-0.18}$ &   $6.99^{+0.32}_{-0.25}$ \\ 
$f_{\rm ellip}$ &   $0.06^{+0.09}_{-0.05}$ &   $0.56^{+0.34}_{-0.39}$ &   $0.23^{+0.15}_{-0.15}$ &   $0.32^{+0.17}_{-0.22}$ &   $0.66^{+0.21}_{-0.27}$ \\ 
$f_{\rm flow}$ &   $0.24^{+0.20}_{-0.17}$ &   $0.65^{+0.24}_{-0.38}$ &   $0.25^{+0.21}_{-0.16}$ &   $0.29^{+0.25}_{-0.19}$ &   $0.31^{+0.31}_{-0.22}$ \\ 
$\theta_e$ (degrees) &   $12.0^{+10.7}_{- 8.3}$ &   $57.2^{+24.9}_{-41.0}$ &   $21.3^{+21.4}_{-14.7}$ &   $47.0^{+16.7}_{-26.5}$ &   $49.7^{+28.8}_{-32.1}$ \\ 
$\sigma_{\rm turb}$ &   $0.008^{+0.028}_{-0.007}$ &   $0.004^{+0.010}_{-0.003}$ &   $0.016^{+0.044}_{-0.013}$ &   $0.013^{+0.036}_{-0.011}$ &   $0.011^{+0.033}_{-0.009}$ \\ 

  \hline
    \end{tabular}
   \medskip
    The definitions of the geometry and dynamical model parameters can be found in Appendix~\ref{appendix_params}.
  \end{minipage}
  \end{table*}

  \begin{table*}
  \caption{Inferred posterior median parameter values and central 68\% credible intervals for $f$ factors of five LAMP 2008 AGNs.  The $f$ factor 
  corresponding to the difference between black hole mass and the virial product measured using the dispersion of the RMS line profile 
  is given as $f_{\sigma}$, while the one corresponding to a virial product measured using the FWHM of the mean line profile is
  given as $f_{\rm FWHM}$. }
  \label{table_f}
  \begin{tabular}{@{}ccc}
   \hline
   Object                     & $\log_{10}(f_\sigma)$  & $\log_{10}(f_{\rm FWHM})$  \\
   \hline   
Arp 151 &   $0.51^{+0.10}_{-0.13}$ &   $-0.24^{+0.10}_{-0.13}$ \\ 
Mrk 1310 &   $1.63^{+0.26}_{-0.27}$ &   $0.79^{+0.26}_{-0.27}$ \\ 
NGC 5548 &   $0.42^{+0.23}_{-0.14}$ &   $-0.58^{+0.23}_{-0.14}$ \\ 
NGC 6814 &   $-0.14^{+0.24}_{-0.18}$ &   $-0.68^{+0.24}_{-0.18}$ \\ 
SBS 1116+583A &   $0.96^{+0.32}_{-0.25}$ &   $0.34^{+0.32}_{-0.25}$ \\ 
   \hline
  \end{tabular}
  \end{table*}

\section{Conclusions}
\label{sect_conclusions}
We have applied direct modeling techniques to a sample of five AGNs from the LAMP 2008 reverberation mapping campaign in order to constrain
the geometry and dynamics of the \Hb\ BLR.  Direct modeling also allows us to measure the black hole mass independently and, by comparison with the virial product
from traditional reverberation mapping analysis, to measure the virial coefficient or $f$ factor for individual AGNs.  We have also measured
the mean $f$ factor for our sample, a number that determines the absolute mass scaling for the whole reverberation mapping sample.  Our main results are as follows:
\begin{enumerate}
 \item The geometry of the BLR is consistent with a thick disk.  The radial distribution of 
 \Hb\ emitting gas is closer to exponential than Gaussian and is viewed closer to face-on than edge-on.  For Arp 151 we find a more detailed
 geometry of a half-cone, where the \Hb\ emission is concentrated towards the outer faces of the disk and the disk mid-plane is mostly opaque, 
 similar to the bowl BLR geometry proposed by \citet{goad12}.
 \item There is preferential \Hb\ emission from the far side of the BLR with respect to the observer, consistent with models where the BLR gas is self-shielding.  
 \item The dynamics of the BLR are consistent with inflowing motions, elliptical orbits, or a combination of both.  Specifically, the dynamics of Arp 151 are
 inferred to be inflowing motions, in agreement with velocity-resolved cross-correlation lag measurements \citep{bentz09} and reconstruction of the transfer function
 using maximum entropy techniques \citep{bentz10}.
 \item The black hole masses for the five objects in our sample are 
 $\log_{10}(M_{\rm BH}/M_\odot) = 6.62^{+0.10}_{-0.13}$ for Arp 151,
$7.42^{+0.26}_{-0.27}$ for Mrk 1310, 
$7.51^{+0.23}_{-0.14}$ for NGC 5548, 
$6.42^{+0.24}_{-0.18}$ for NGC 6814, 
and $6.99^{+0.32}_{-0.25}$ for SBS 1116+583A.
 \item Using our measurements of the black hole mass and virial products based on the dispersion of the RMS line profile, we measure
 the $f$ factors for the AGNs in our sample to be 
  $\log_{10}(f_{\sigma}) = 0.51^{+0.10}_{-0.13}$ for Arp 151,
$1.63^{+0.26}_{-0.27}$ for Mrk 1310, 
$0.42^{+0.23}_{-0.14}$ for NGC 5548, 
$-0.14^{+0.24}_{-0.18}$ for NGC 6814, 
and $0.96^{+0.32}_{-0.25}$ for SBS 1116+583A.
 Using instead the virial products based on the FWHM of the mean line profile, we find that 
  $\log_{10}(f_{\rm FWHM}) = -0.24^{+0.10}_{-0.13}$ for Arp 151,
$0.79^{+0.26}_{-0.27}$ for Mrk 1310, 
$-0.58^{+0.23}_{-0.14}$ for NGC 5548, 
$-0.68^{+0.24}_{-0.18}$ for NGC 6814, 
and $0.34^{+0.32}_{-0.25}$ for SBS 1116+583A.  
 \item The $f$ factors for individual AGNs are correlated with inclination angle, but not with black hole mass, AGN optical luminosity, or Eddington ratio.  
 \item Neither the $f$ factors nor the inclination angles for individual AGNs are strongly correlated with the ratio of the FWHM to the line dispersion 
 in the mean \Hb\ spectrum, as would be expected if line shape correlated strongly with viewing angle of the BLR.  However, we do find a tentative correlation between
 the ratio of the FWHM to the line dispersion and black hole mass.
 \item By combining the posterior distributions of $f$ for each AGN, we measure mean values of $f$ for the sample.  With virial products based on the dispersion of the 
 RMS line profile, we measure a mean value of $\log_{10}(f_\sigma)$ of
 $0.68 \pm 0.40$ with a dispersion in $\log_{10}(f_\sigma)$ of $0.75 \pm 0.40$, 
 and using virial products based on the FWHM of the mean line profile we measure
a mean $\log_{10}(f_{\rm FWHM})$ value of 
$-0.07 \pm 0.40$ with a dispersion in $\log_{10}(f_{\rm FWHM})$ of $0.77 \pm 0.38$.  
These values of the mean $f$ factor are meant to illustrate the capabilities of the direct modeling approach and should not 
be used to calibrate black hole masses from reverberation mapping until the sample size is larger and more representative of the overall AGN population. 
\end{enumerate}

The modeling results presented here demonstrate the capabilities of the direct modeling approach and show that significant information about the BLR 
geometry and dynamics is encoded in high-quality reverberation mapping datasets.  We find that the five AGNs in our sample have similar geometric 
and kinematic features, suggesting that the BLR may also be similar in other Seyfert 1 galaxies with low luminosities, black hole masses of $10^{6.5-7.5}M_\odot$, and small Eddington ratios.
By applying the direct modeling approach to a larger sample of AGNs, we can determine if and how the properties of the BLR might change with increasing luminosity, accretion rate, and black hole mass.  

Our results also demonstrate the feasibility of measuring black hole masses independently of the $f$ factor in Seyfert 1 galaxies.  For the reverberation mapping datasets
shown here, black hole masses can be constrained to $0.15-0.3$ dex uncertainty depending upon data quality and degeneracy of the black hole mass
with the geometrical properties of the BLR, such as inclination angle of the observer and opening angle of the disk.  In addition, the BLR kinematics inferred
for our sample are consistent with bound orbits, suggesting that the \Hb-emitting BLR is not significantly affected by disk winds or outflows.  
This is an important consistency check for reverberation mapped black hole masses because they are measured by assuming the BLR gas orbits are 
dominated by the gravity of the black hole.
Future versions of our BLR model will explore the issue of non-gravitational forces further and 
relate broad line emission to the properties of the emitting gas
by incorporating the results of photoionization physics.

\section*{Acknowledgements}  
AP would like to thank Jason Kaufman for helpful discussions. 
AP acknowledges support from the NSF through the Graduate Research Fellowship Program. 
AP, BJB, and TT acknowledge support from the Packard Foundation through a Packard Fellowship to TT and support from the NSF through awards NSF-CAREER-0642621 and NSF-1108835.    
BJB is partially supported by the Marsden Fund (Royal Society of New Zealand). 
Research by AJB is supported by NSF grant AST-1108835.
MCB gratefully acknowledges support from the NSF through CAREER Grant AST-1253702.
JHW acknowledge support by Basic Science Research Program through the National Research Foundation of Korea funded by the Ministry of Education, Science and Technology (2012-006087).
The LAMP 2008 project was also supported by NSF grants AST-0548198 (UC Irvine) and AST-0507450 (UC Riverside).

\appendix
\onecolumn
\section{Definition of Model Parameters}
\label{appendix_params}

\subsection{Geometry Model Parameters}
We use a Gamma distribution to model the radial distribution of point particles in the BLR:
\begin{eqnarray}
 p(x|\alpha,\theta) \propto x^{\alpha-1} \exp \left( -\frac{x}{\theta} \right).
\end{eqnarray}
We then allow the Gamma distribution to be offset from the origin by an amount $r_{\rm min}$
plus the Schwarzschild radius, $R_s = 2GM/c^2$,
 and perform a change of variables between $(\alpha,\,\theta)$ and $(\mu,\,\beta,\,F)$
such that
\begin{eqnarray}
 \mu &=& R_s + r_{\rm min} + \alpha\theta \label{eqn_mu} \\
 \beta &=& \frac{1}{\sqrt{\alpha}} \label{eqn_beta}\\
 F &=& \frac{r_{\rm min}}{r_{\rm min}+\alpha\theta} \label{eqn_f}
 \end{eqnarray}
where $\mu$ is the mean radius, $\beta$ determines
the shape of the Gamma distribution, and $F$ is the 
fraction of $\mu$ corresponding to $r_{\rm min}$.  
The prior on $\mu$ is uniform in the log of the 
parameter between $1.02\times10^{-3}$ light days and the time span 
between the first and last measurement of the continuum or line flux,
while the prior on $\beta$ is uniform between 0 and 2
and the prior on $F$ is uniform between 0 and 1.
The standard deviation for the radial distribution is given by 
$\sigma_r = \mu \beta (1 - F)$.
We can also calculate the numerical mean radius $r_{\rm mean}$, the numerical mean time lag $\tau_{\rm mean}$, 
 and the numerical median time lag $\tau_{\rm median}$ for a specific realization of 
point particle positions.
The direct modeling
results in Table~\ref{table_results} include values for $r_{\rm mean}$,
$r_{\rm min}$, $\sigma_r$, $\tau$, and $\beta$.
The geometry of the BLR is further defined by $\theta_o$, the half-opening angle of the BLR disk.  Values of $\theta_o\to 0$ (90) degrees correspond
to thin disk (spherical) geometries and the prior is uniform between 0 and 90 degrees. 
The inclination angle, $\theta_i$, is the angle by which an observer views the BLR.  Values of $\theta_i \to 0$ (90) degrees correspond
to face-on (edge-on) geometries and the prior is uniform in the cosine of the inclination angle between 0 and 90 degrees. 
We weight the emission from each point particle by a cosine function:
\begin{eqnarray}
 W(\phi) = \frac{1}{2} + \kappa \cos \phi. \label{eqn_kappa}
\end{eqnarray}
where $W$ is the weight (between 0 and 1) given to each point particle, $\phi$ is the angle between the observer's line of sight
to the central source and the point particle's line of sight to the central source, and $\kappa$ is a
parameter with uniform prior between $-0.5$ and 0.5.  Values of $\kappa \to -0.5$ (0.5) correspond to the far (near) side
of the BLR producing more line emission.
We also include the option for preferential emission from the faces of the BLR disk by 
changing the angle $\theta$ for a point particle's displacement from a flat to thick disk, given by
\begin{eqnarray}
 \theta = {\rm acos} \left( \cos \theta_o + (1 - \cos \theta_o)\times U^\gamma \right)  \label{eqn_gamma}
\end{eqnarray}
where $U$ is a random number drawn uniformly between the values of 0 and 1.  Values of 
$\gamma \to 1$ (5) correspond to uniform concentrations of point particles in the disk (more point particles
along the faces of the disk), where $\gamma$ has a uniform prior between 1 and 5.
Finally, we allow the midplane of the BLR to range between opaque and transparent, where 
$\xi$ is the fraction of the point particles below the midplane that are not moved to the top half.
For $\xi \to 1$ (0) the midplane is transparent (opaque), where $\xi$ has a uniform prior between 
0 and 1. 

\subsection{Dynamical Model Parameters}
The dynamics of the BLR are determined by the black hole mass, $M_{\rm BH}$, which
has a uniform prior in the log of the parameter between $2.78\times10^4$ and $1.67\times10^9 M_\odot$.
We draw the velocities for the point particles from two distributions in the plane of radial and tangential velocities.  
The fraction of point particles with velocities drawn from the distribution centered around the circular
orbit value is given by $f_{\rm ellip}$, which has a uniform prior between 0 and 1.  
The remaining point particles have velocities drawn
from the distribution centered around either the radial 
inflowing or outflowing escape velocity values, where $0 < f_{\rm flow} < 0.5$ corresponds
to the inflowing distribution and $0.5 < f_{\rm flow} < 1$ corresponds
to the outflowing distribution, and where $f_{\rm flow}$ has a uniform prior between 0 and 1.  
The inflow/outflow-centered distributions can also
be rotated by an angle $\theta_{e}$ towards the circular orbit-centered distribution, 
where $\theta_{e}$ has a uniform prior between 0 and 90 degrees.
Finally, we include additional macroturbulent velocities given by:
\begin{eqnarray}
 v_{\rm turb} = \mathcal{N}(0,\sigma_{\rm turb}) |v_{\rm circ}|
\end{eqnarray}
where $v_{\rm circ}$ is the circular orbit velocity and $\sigma_{\rm turb}$ is the 
standard deviation of the Gaussian distribution from which a random 
macroturbulent velocity component is drawn.  
 $\sigma_{\rm turb}$ has a uniform prior in the log of the parameter
between 0.001 and 0.1.

\section{Calculating the Mean $\lowercase{f}$ Factor}
\label{appendix_f}

For each of the five AGNs in our sample, we can compute the posterior distribution
for the $f$ factor that relates the black hole mass to either the velocity
dispersion or the FWHM of the broad emission line. Here we describe the method
used to constrain the distribution of $f$ values from the modelling results \citep[see][for examples using the same approach]{hogg10, brewer14}.
Consider a collection of $N$ objects, each of which has a property $\theta$ which
we infer from data $D$. Modelling each object yields a posterior distribution
\begin{eqnarray}
p(\theta_i | D_i) \propto \pi(\theta_i)p(D_i | \theta_i)
\end{eqnarray}
where $\pi(\theta_i)$ is the prior used in the modelling, which is the same
for each object. In practice, since we are using MCMC, the posterior
distributions $p(\theta_i | D_i)$ are represented computationally by samples.
In our particular application, $\theta \equiv \log_{10}(f)$.

Unfortunately, the use of the $\pi(\theta_i)$ prior for each object
implies we do not expect the
objects to be clustered around a typical $\theta$ value. If we did expect such
clustering, we should have used a different prior for the $\{\theta_i\}$, such
as a normal distribution:
\begin{eqnarray}
p(\{\theta_i\} | \mu_\theta, \sigma_\theta) &=& 
\prod_{i=1}^N \frac{1}{\sigma_\theta\sqrt{2\pi}}
\exp\left[
-\frac{1}{2\sigma_\theta^2}\left(\theta_i - \mu_\theta\right)^2
\right]
\end{eqnarray}
This is the prior, conditional on two new hyperparameters describing the
typical value $\mu_\theta$ that the objects are clustered around, and the
scatter $\sigma_\theta$. To complete the inference we also need to assign
a prior to $\mu_\theta$ and $\sigma_\theta$, which we take to be vague.
Using this model, we can summarise our uncertainty about the properties of the
sample by calculating the posterior distribution for $\mu_\theta$ and
$\sigma_\theta$. Alternatively the posterior distribution for the {\it actual}
mean $\frac{1}{N}\sum_{i=1}^N \theta_i$ could be calculated, but the former
approach allows for generalisation beyond the current sample.

The posterior distribution for the hyperparameters is
\begin{eqnarray}
p(\mu_\theta, \sigma_\theta | \{D_i\})
&\propto&
p(\mu_\theta, \sigma_\theta )p(\{D_i\} | \mu_\theta, \sigma_\theta)\\
&\propto&
p(\mu_\theta, \sigma_\theta ) \int \prod_{i=1}^Np(\theta_i, D_i | \mu_\theta, \sigma_\theta) \, d^N\theta_i\\
&\propto&
p(\mu_\theta, \sigma_\theta ) \int \prod_{i=1}^Np(\theta_i | \mu_\theta, \sigma_\theta)p(D_i | \theta_i, \mu_\theta, \sigma_\theta) \, d^N\theta_i\\
&\propto&
p(\mu_\theta, \sigma_\theta ) \int \prod_{i=1}^Np(\theta_i | \mu_\theta, \sigma_\theta)p(D_i | \theta_i)\, d^N\theta_i\\
&\propto&
p(\mu_\theta, \sigma_\theta ) \int \prod_{i=1}^Np(\theta_i | \mu_\theta, \sigma_\theta)\frac{\pi(\theta_i)}{\pi(\theta_i)}p(D_i | \theta_i)\, d^N\theta_i\\
&\propto&
p(\mu_\theta, \sigma_\theta ) \prod_{i=1}^N \left< \frac{p(\theta_i | \mu_\theta, \sigma_\theta)}{\pi(\theta_i)}\right>.\label{eq:expectation}
\end{eqnarray}
where the expectation is taken with respect to the posterior distributions we
have actually sampled, and can be computed straightforwardly. Essentially,
Eq~\ref{eq:expectation} favors $(\mu_\theta, \sigma_\theta)$ values that
place a lot of probability in regions that overlap with the posteriors that
we found.

\end{document}